\documentclass[12pt, draftclsnofoot, onecolumn]{IEEEtran}
 \usepackage{amsmath,amssymb}
 \usepackage{subfigure}
 \usepackage{graphicx,graphics,color,psfrag}
 \usepackage{cite,balance}
 \usepackage{caption}
 \allowdisplaybreaks
 \usepackage{algorithm}
 \usepackage{accents}
 \usepackage{amsthm}
 \usepackage{bm}
 \usepackage{algorithmic}
 \usepackage[english]{babel}
 \usepackage{multirow}
 \usepackage{enumerate}
 \usepackage{cases}
 \usepackage{stfloats}
 \usepackage{dsfont}
 \usepackage{color,soul}
 \usepackage{amsfonts}
 \usepackage{cite,graphicx,amsmath,amssymb}
 \usepackage{subfigure}
 \usepackage{fancyhdr}
 \usepackage{hhline}
 \usepackage{graphicx,graphics}
 \usepackage{array,color}
 \usepackage{amsmath}
 \usepackage{booktabs}

\newtheorem{lemma}{Lemma}

\newtheorem{proposition}{Proposition}

\newtheorem{remark}{\bf Remark}

\def\qed{$\Box$}

\def\proof{\noindent{\emph{Proof:} }}

\def
\endproof{\hspace*{\fill}~\qed
\par
\endtrivlist\unskip}
\def\endproof{\hspace*{\fill}~\qed\par\endtrivlist\vskip3pt}

\def\phi{\varphi}

\def\l{\left}
\def\r{\right}
\def\({\left(}
\def\){\right)}

\def\bx{{\mathbf{x}}}
\def\by{{\mathbf{y}}}

\def\b0{{\mathbf{0}}}

\begin{document}
\setlength{\topskip}{-3pt}

\title{\huge Integrated Sensing, Communication, and Compu- tation Over-the-Air: MIMO Beamforming Design}
\author{Xiaoyang Li, Fan Liu, Ziqin Zhou, Guangxu Zhu, Shuai Wang, Kaibin Huang, and Yi Gong 
\thanks{Xiaoyang Li, Fan Liu, Ziqin Zhou, and Yi Gong are with the the Department of Electrical and Electronic Engineering
(EEE), Southern University of Science and Technology (SUSTech), Shenzhen, China. Guangxu Zhu is with the Shenzhen Research Institute of Big Data, Shenzhen, China. Shuai Wang is with the Shenzhen Institute of Advanced Technology, Chinese Academy of Sciences, Shenzhen, China. Kaibin Huang is with the Department of EEE, The University of Hong Kong, Hong Kong. Corresponding author: Yi Gong (e-mail: gongy@sustech.edu.cn).} 
}
\maketitle

\vspace{-15mm}
\begin{abstract}
To support the unprecedented growth of the Internet of Things (IoT) applications, tremendous data need to be collected by the IoT devices and delivered to the server for further computation. By utilizing the same signals for both radar sensing and data communication, the \emph{integrated sensing and communication} (ISAC) technique has broken the barriers between data collection and delivery in the physical layer. By exploiting the analog-wave addition in a multi-access channel, \emph{over-the-air computation} (AirComp) enables function computation via transmissions in the physical layer. The promising performance of ISAC and AirComp motivates the current work on developing a framework called \emph{integrated sensing, communication, and computation over-the-air} (ISCCO). Two schemes are designed to support \emph{multiple-input-multiple-output} (MIMO) ISCCO simultaneously, namely the separated and shared schemes. The separated scheme splits antenna array for radar sensing and AirComp respectively, while all the antennas transmit a joint waveform for both radar sensing and AirComp in the shared scheme. The performance metrics of radar sensing and AirComp are evaluated by the mean squared errors of the estimated target response matrix and the received computation results, respectively. The design challenge of MIMO ISCCO lies in the joint optimization of beamformers for sensing, communication, and computation at both the IoT devices and the server, which results in a non-convex problem. To solve this problem, an algorithmic solution based on the technique of semidefinite relaxation is proposed. The results reveal that the beamformer at each sensor needs to account for supporting dual-functional signals in the shared scheme, while dedicated beamformers for sensing and AirComp are needed to mitigate the mutual interference between the two functionalities in the separated scheme. The use case of target location estimation based on ISCCO is demonstrated in simulation to show the performance superiority.
\end{abstract}



\section{Introduction}
The next-generation wireless networks (6G and beyond) have been envisioned as a vital enabler for the emerging \emph{Internet of Things} (IoT) services, such as autonomous vehicles, extended reality, artificial intelligence, smart cities, and digital twins \cite{saad2019vision}. To support the prosperous applications, tremendous data need to be collected from the environment by the IoT devices and delivered to the server for further fusion and computation \cite{cui2021integrating}. In conventional data processing pipelines, the above operations are individually accomplished with limited mutual assistance and rare integration \cite{liu2020joint}. By utilizing the same spectrum and signals for both radar sensing and data communication, the \emph{integrated sensing and communication} (ISAC) technique has broken the barriers between data collection and delivery in the physical layer \cite{liu2021integrated}. Nevertheless, the operation of computation is still isolated as it mainly lies in the upper layers.

For reducing the overheads and improving the efficiency, it is natural to integrate the operations of sensing, communication, and computation together. By exploiting the analog-wave addition in a multi-access channel, \emph{over-the-air computation} (AirComp) realizes fast wireless data fusion via simultaneous transmissions in the physical layer \cite{zhu2021over}. Prompted by AirComp, the three operations are expected to be unified within a single signal transmission, which motivates the current work on developing a framework called \emph{integrated sensing, communication, and computation over-the-air} (ISCCO).


In the ISCCO framework, an IoT system is considered for supporting sensing, communication, and computation simultaneously. To be specific, multiple multi-antenna IoT sensors transmit radar signals to detect a target and data symbols to a multi-antenna server for data fusion via AirComp. The dual-functional radar sensing and AirComp can be achieved by two schemes, namely \emph{shared scheme} and \emph{separated scheme}. In the shared scheme, the whole antenna array at each sensor are exploited for transceiving a signal both serving as a radar pulse and a data carrier. In the separated scheme, the antennas at each sensor are divided into two groups for radar sensing and data transmission. As the \emph{key performance indicators} (KPIs) for radar sensing and AirComp are the \emph{mean squared errors} (MSE) of target estimation and function computation, respectively, there exists a natural tradeoff between the performance of the two functionalities, which is reflected in the beamforming design. This introduces \emph{coupling between sensing and AirComp}, and hence necessitates the joint design of radar signal beamforming, data transmission beamforming, and data aggregation beamforming. The beamforming designs together with the performance analysis for both the shared and separated schemes are investigated in this paper.

The contributions of this work are summarized as follows.

\begin{itemize}
\item {\bf Transmission and Aggregation Beamforming Design in the Shared Scheme}: As the signal transmitted by each sensor in the shared scheme serves both as the radar probing pulse and data carrier, only one beamformer needs to be designed at each sensor, which is known as the transmission beamformer. The radar sensing target can be extracted from the statistic information of the reflected signal via \emph{maximum likelihood estimation} (MLE). As for AirComp, a beamformer at the server is deployed for equalizing the received signals, which is known as the data aggregation beamformer. The joint transmission and data aggregation beamforming design is formulated as a semidefinite programming problem for minimizing the computation error in AirComp under the constraints of radar sensing requirement and power budget for each sensor. The solving approach based on semidefinite relaxation is applied to obtain the desired design.

\item {\bf Radar and Communication Beamforming Design in the Separated Scheme}: As there are two signals transmitted by each sensor with one for radar sensing and another for data transmission, two corresponding beamformers are needed to be designed at each sensor. Moreover, the existence of radar signal will exacerbate the interference at the server, which results in a more complex performance metrics of AirComp. The coupling between radar sensing beamformer, data transmission beamformer, and data aggregation beamformer makes the optimization problem even challenging to be coped with. To tackle such a problem, an orthogonal constraint for radar sensing beamforming is imposed in the beamforming design.

\item {\bf Comparison between Shared and Separated Schemes}: Simulation is conducted to compare the performance of the shared and separated schemes. Particularly, the beamformers in the shared scheme try to mitigate the noise in AirComp while guaranteeing the sensing accuracy. In the separated scheme, the beamformers are further accounted for reducing the interference of radar signals on AirComp. The radar sensing MSE in the separated scheme solely depends on the maximum MSE tolerance for reducing the AirComp error, while the sensing MSE in the shared scheme is relevant to multiple parameters such as the number of antennas.

\item {\bf Target Location Estimation based on ISCCO}: To illustrate the performance of ISCCO design, the use case of target location estimation is conducted. Specifically, the target location is estimated locally by each sensor based on the information extracted from its received radar signals and transmitted to the server via AirComp. The averaged estimated target location received by the server is compared with the ground truth.
\end{itemize}

The remainder of the paper is organized as follows. Section II reviews the state-of-the-art techniques of AirComp and ISAC. Section III introduces the ISCCO system model. Sections IV and V present the shared and separated beamforming designs for achieving the best ISCCO performance, respectively. Section VI further illustrates the performance of target location estimation based on ISCCO. Simulation results are provided in Section VII, followed by concluding remarks in Section VIII.


\section{Background of AirComp and ISAC}

\subsection{AirComp}
The idea of AirComp can be traced back to the pioneering work studying function computation in sensor networks \cite{nazer2007computation}, where the distributed sensing values are analogly modulated and transmitted over a \emph{multi-access channel} for reliable function computation at a server. The importance of the work lies in the counter-intuitive finding that interference caused by simultaneous transmission can be exploited to facilitate computation. The transmission synchronization over sensors was further investigated in \cite{katabi2015airshare}. 


Driven by the need of fast data aggregation in IoT, AirComp has been applied to supporting function computation via data transmission from multiple sensors to the server. The functions that can be calculated by AirComp has the general format $h(\cdot)$ as shown below:
\begin{align}\label{AirCompfunc}
\by=h(\bx_1, \bx_2, \cdots, \bx_K) = f\l(\sum_{k=1}^{K}g_{k}(\bx_k)\r),
\end{align}
where $\{\bx_k\}$ represents the distributed data samples, $f(\cdot)$ and $g_{k}(\cdot)$ represent post-processing at the server and pre-processing at a device, respectively. The summation in \eqref{AirCompfunc} is achieved by simultaneous analog transmission to exploit the wave-addition of the multi-access channel. Consequently, the function computation is performed “over-the-air” and the result is directly received by the server. The class of functions having the above form is known as \emph{nomographic functions} such as averaging and geometric mean. Typical functions in this class are summarized in Table \ref{summary:table1}. Simultaneous transmission in AirComp achieves low latency independent of the number of devices and saves the spectrum resources.

\begin{table}[t]
\centering
\caption{Examples of nomographic functions in AirComp.}
\begin{tabular}{|p{4cm}|p{4cm}|}
\hline
\bf{Name} & \bf{Expression} \\
\hline
Arithmetic Mean &  $y = \frac{1}{K}\sum_{k=1}^K x_k$ \\ 
\hline
Weighted Sum &  $y = \sum_{k=1}^K \omega_k x_k$ \\ 
\hline
Geometric Mean &  $y = \l(\prod_{k=1}^K x_k \r)^{1/K}$ \\ 
\hline
Polynomial &  $y = \sum_{k=1}^K \omega_k x_k^{\beta_k}$ \\ 
\hline
Euclidean Norm &  $y = \sqrt{\sum_{k=1}^K x_k^2}$ \\ 
\hline
\end{tabular}
\label{summary:table1}
\end{table}

To accelerate the computation of multiple functions, the features of \emph{multi-modal sensing} as well as the prevalence of antenna arrays at both servers and devices were exploited to enable the \emph{multiple input multiple output} (MIMO) AirComp \cite{chen2018over}. In MIMO AirComp, the spatial \emph{degrees-of-freedom} is leveraged to spatially multiplex multi-function computation simultaneously and reduce computation errors by noise suppression \cite{zhu2018mimo}. Along this vein, the MIMO AirComp was integrated with the wireless power transfer technique to achieve self-sustainable AirComp for low-power devices \cite{li2019wirelessly}. The reduced-dimension design of AirComp was investigated in \cite{wen2019reduced} for clustered IoT networks. To overcome the reliance on \emph{channel station information} (CSI), a blind MIMO AirComp technique without requiring CSI access was proposed in \cite{dong2020blind} for low-complexity and low-latency IoT networks. Facing the practical scenario with fading channels, the optimal power control was designed in \cite{cao2020optimized} to deal with the channel distortion, while the tradeoff between the computation effectiveness and the energy efficiency was analyzed in \cite{liu2020over}. The hybrid beamforming for massive MIMO AirComp was studied in \cite{zhai2021hybrid}. 

Due to its promising performance in fast function computation, AirComp has been deployed in a series of IoT applications, including \emph{federated edge learning} (FEEL) \cite{zhu2019broadband,sun2021dynamic,amiri2020federated,yang2020federated,guo2020analog,xu2021learning,zhang2021gradient,zhu2020one,liu2020privacy}, \emph{reconfigurable intelligent surface} (RIS) assisted communication \cite{ni2021federated}, \emph{unmanned aerial vehicle} (UAV) communication \cite{fu2021uav}, autonomous driving \cite{wang2021edge}, and MapReduce over the edge cloud network \cite{han2021over}. As for FEEL, the wave-addition of AirComp is in perfect match with the aggregation of local training results at the server for global model updating, which has attracted great research efforts. The existing literature have investigated the implementation of AirComp in FEEL from different perspectives, including communication-learning tradeoff \cite{zhu2019broadband}, devices scheduling \cite{sun2021dynamic}, update compression \cite{amiri2020federated}, beamforming design \cite{yang2020federated}, hyper-parameters control \cite{guo2020analog}, learning rate control \cite{xu2021learning}, power control \cite{zhang2021gradient}, digital modulation \cite{zhu2020one}, and data privacy \cite{liu2020privacy}.

Despite the wide applications of AirComp in learning and communication systems, the incorporation of data sensing remains as an uncharted area in AirComp, which deserves to be investigated in this paper. 

\subsection{ISAC}
The origin of ISAC can be traced back to the early work in joint radar-communication, where information was embedded into a group of radar pulses \cite{mealey1963method}. In practice, the S-band (2-4 GHz) and C-band (4-8 GHz) occupied by radar applications might be shared with communication systems \cite{chapin2012shared}. Consequently, a series of treatises focuses on investigating the co-existence of radar and communications systems. Particularly, an opportunistic spectrum sharing scheme was proposed in \cite{saruthirathanaworakun2012opportunistic}, where the communication signals are sent when the spectrum is not occupied by radar. Despite its easy implementation, the radar and communication functions are unable to work simultaneously. To overcome such a drawback, a null-space projection method was carried out to support the co-existence of MIMO radar sensing and communication \cite{sodagari2012projection}, where the radar signals are projected onto the null-space of the interference channel for the communication link. Nevertheless, such projection might harm the optimality of radar signal beamforming and thus results in performance loss for the radar sensing. 

Research has been conducted on improving the performance of radar sensing and communication based on convex optimization techniques. In \cite{li2016mimo}, the designs of radar beamformer and communication covariance matrix were jointly optimized to maximize the radar sensing \emph{signal to interference plus noise ratio} (SINR) subject to specific capacity and power constraints. As for co-existence of MIMO radar and \emph{multi-user MIMO} (MU-MIMO) communications, a robust beamforming design with imperfect CSI was proposed in \cite{liu2017robust}, where the radar sensing accuracy is maximized under the SINR requirements by communication and the power budget. Moreover, the multi-user interference was exploited as a source of transmission power in \cite{liu2018mimo}, based on which a novel beamforming design was proposed. In order to develop an optimal communication reception strategy in the presence of the radar interference, a communication receiver was designed in \cite{zheng2017adaptive} to demodulate the communication data while removing the radar interference iteratively using a \emph{successive interference cancellation} (SIC) algorithm. It should be noted that the side-information including CSI, radar probing waveforms, and communication modulation formats need to be frequently exchanged between the radar and communication devices to support the coexistence. Though such cooperation might be achieved by deploying a control center connecting both systems via a wireless link or a backhaul channel, the implementation will impose extra complexity on the system \cite{li2017joint}.

To reduce the side-information exchange overheads, an advanced co-existence scheme was proposed in \cite{paul2016survey}, where a dual-functional system supporting both radar and communications was designed. From the perspective of information theory, the performance of radar and communications were unified based on the rate distortion theory \cite{chiriyath2015inner}. The implementation in practice was conducted by the dual-functional waveform design, which supports target detection as well as data transmission simultaneously \cite{blunt2011performance}. Along this vein, the integrated radar and communication waveform was designed in single antenna systems \cite{moghaddasi2016multifunctional}. As a step forward, the work of \cite{hassanien2015dual} brought the integrated waveform design into the MIMO systems, where the information bits are embedded in the sidelobe of the radar transmitting beampattern. Accounting for the multi-user communication system, a series of transmitting beamforming designs were carried out in \cite{liu2018mu} w.r.t. both shared and separated schemes. Aiming at reducing signal distortion, the constant modulus waveforms was further conceived in the dual-functional beamforming design \cite{liu2018toward}. 

The benefit of spectrum sharing makes ISAC a popular technology that has beed applied in a series of systems, such as millimeter-wave radar and communication networks \cite{kumari2017ieee}, RIS systems \cite{wang2021joint}, smart home \cite{huang2020joint}, edge learning systems \cite{zhang2021accelerating}, vehicular networks \cite{yuan2020bayesian}, and UAV systems \cite{lyu2021joint}. Particularly, an IEEE 802.11ad-based radar was deployed in millimeter-wave band for supporting an automotive radar and communication network \cite{kumari2017ieee}. To mitigate the multi-user interference in ISAC, the joint waveform and discrete phase shift design was carried out by employing RIS \cite{wang2021joint}. As for smart home, the traditional sensing devices were empowered with communication capability, while the sensing capability of WiFi signals were enhanced \cite{huang2020joint}. ISAC was further applied to accelerate the edge learning process by designing wireless signals for the dual purposes of dataset generation and uploading. In vehicular networks, the wireless sensing functionality was exploited to acquire vehicles’ states and facilitate the communication \cite{yuan2020bayesian}. In UAV-enabled ISAC system, the maneuver and beamforming designs were jointly optimized to communicate with multiple users and sense potential targets simultaneously \cite{lyu2021joint}.

Among the rich literature on ISAC, the operation of data computation is always overlooked as it lies in the upper layers. Fortunately, AirComp enables fast function computation via transmissions in the physical layer. Therefore, it is natural to integrate the operations of sensing, communication, and computation via the combination of ISAC and AirComp.

\section{System Model}
The considered MIMO ISCCO system comprises one common target, one \emph{access point} (AP) equipped with $N_a$ antennas, and $M = |\mathcal{M}|$ radar sensors equipped with $N_s$ antennas. Each radar sensor can simultaneously transmit probing signals to detect the target and data symbols to the AP for AirComp. The ISCCO phase is divided into $T$ time slots. The operations of different sensors are synchronized using a reference clock broadcast by the server (see e.g., \cite{katabi2015airshare}). The CSI between the AP and each sensor is estimated individually at each sensor from broadcasted pilot signals and then passed to the AP subsequently. For simplicity, channels are assumed to vary following the \emph{block-fading} model. In other words, each channel remains fixed within a phase and varies over different phases. The maximum transmit power of each sensor is $P$. Two ISCCO schemes are considered in this paper, namely the shared scheme and the separated scheme. The notations are summarized in Table \ref{summary:table2}.

\begin{table*}[t]
\centering
\caption{Notation.}
\begin{tabular}{|p{1cm}|p{5.5cm}|p{1cm}|p{7cm}|}
\hline
\bf{Symbol} & \bf{Definition} & \bf{Symbol} & \bf{Definition}\\
\hline
$M$ &  Number of sensors & $\bold{G}_{im}$ & Target response matrix between sensor $i$ and $m$ \\ 
\hline
$K$ & Number of AirComp functions & $\bold{H}_{m}$ & Data transmission channel from sensor $m$ to AP \\ 
\hline
$N_a$ &  Number of antennas at the AP & $\bold{Q}_{im}$ & Direct radar channel between sensor $i$ and $m$\\ 
\hline
$N_s$ &  Number of antennas at each sensor & $\bold{R}_{m}$ & Radar signal channel from sensor $m$ to AP \\ 
\hline
$\bold{n}_c$ & Noise of data transmission channel & $\eta_m$ & Sensing MSE threshold of sensor $m$ \\ 
\hline
$T$ & Number of time slots in ISCCO phase & $\bold{C}_{im}$ & Data reflection channel between sensor $i$ and $m$\\ 
\hline
$P$ & Power budget of each sensor & $\bold{O}_{im}$ & Direct data channel between sensor $i$ and $m$\\ 
\hline
$\bold{s}_m[t]$ &  Radar signals of sensor $m$ at time $t$ & $\bold{\Omega}_{m}[t]$ & Interference signal for sensor $m$ at time $t$\\ 
\hline
$\bold{d}_m[t]$ & Data symbols of sensor $m$ at time $t$ & $\bold{\hat{Y}}_m$ & Sufficient statistic matrix of sensor $m$\\ 
\hline
$\bold{x}_m[t]$ & Signals transmitted by sensor $m$ at time $t$ & $\bold{y}_m[t]$ & Target reflection signal of sensor $m$ at time $t$\\ 
\hline
$\bold{n}_r$ & Noise of radar signal channel & $N_c$ & Number of antennas at sensor for data transmission\\ 
\hline
$\bold{W}_m$ & Data transmission beamformer of sensor $m$ & $N_r$ & Number of antennas at sensor for radar sensing\\ 
\hline
$\bold{F}_m$ & Radar signal beamformer of sensor $m$ & $N_{tx}$ & Number of radar transmitting antennas at sensor\\ 
\hline
$\bold{A}$ & Data aggregation beamformer at AP & $N_{rx}$ & Number of radar receiving antennas at sensor\\ 
\hline
\end{tabular}
\label{summary:table2}
\end{table*}

\begin{figure*}[ht]
\centering
\includegraphics[scale=0.45]{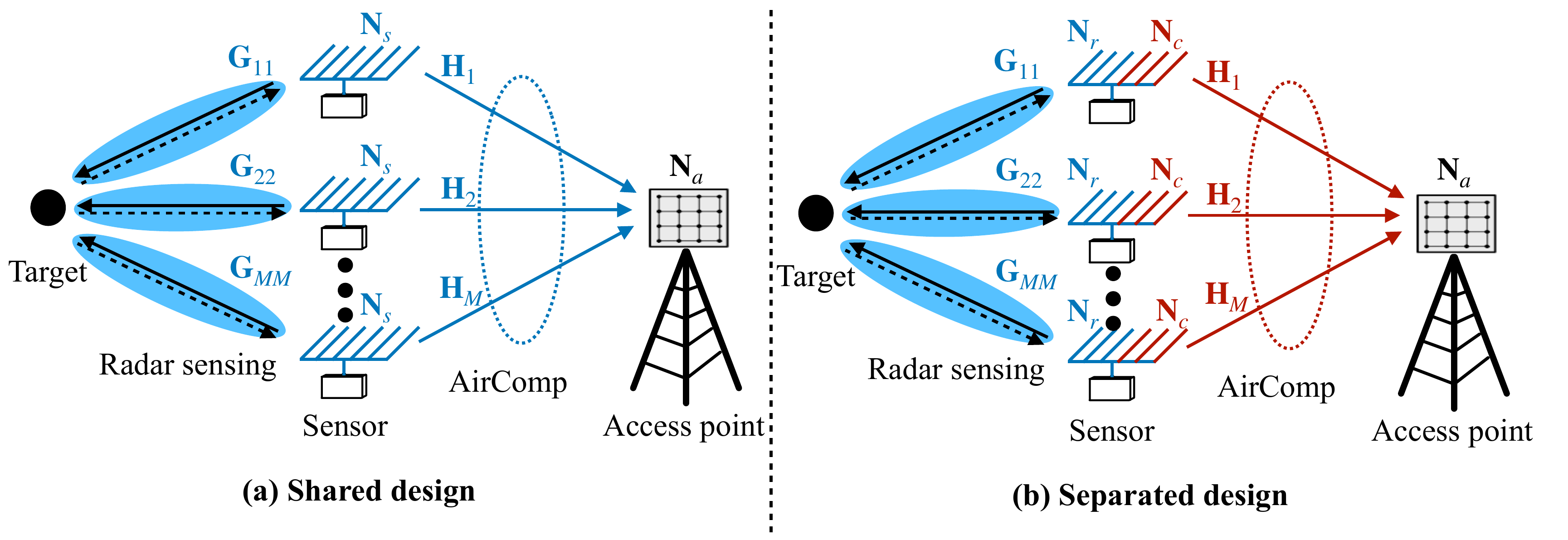}
\caption{Integrated sensing and AirComp system.}
\label{FigSys}
\end{figure*}

\subsection{Shared Scheme}
\begin{figure*}[ht]
\centering
\includegraphics[scale=0.5]{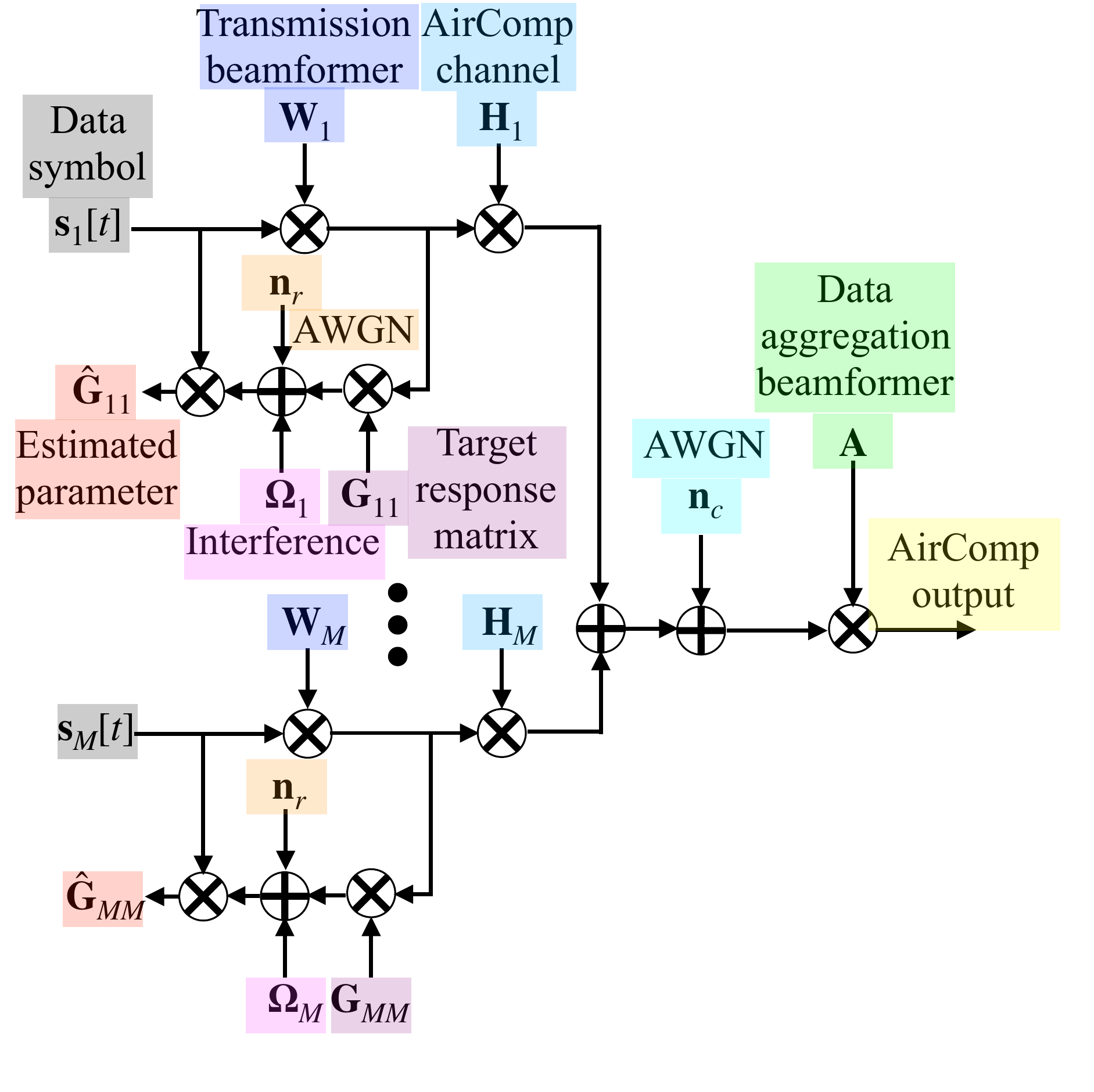}
\caption{Diagram of shared scheme.}
\label{FigShare}
\end{figure*}

As shown in Fig. \ref{FigSys} (a), all the $N_s = N_{tx} + N_{rx}$ antennas at each sensor are shared for both radar sensing and data transmission, where $N_{tx}$ antennas are for signal transmitting and $N_{rx}$ for receiving. The signals transmitted by the sensors have dual functions for both sensing the target and carrying the data symbols to the AP simultaneously. The data symbols to be transmitted by the $m$-th sensor at the $t$-th slot can be expressed as a vector denoted by $\bold{s}_m[t] \in \mathbb{C}^{K \times 1}$, where $K$ represents the number of functions to be calculated via AirComp. The data symbols are assumed to be i.i.d. among different sensors and functions with zero mean and unit variance, i.e., $\mathbb{E}_t[\bold{s}_m[t] \bold{s}_m^H[t]] =  \bold{I}$ and $\mathbb{E}_t[\bold{s}_m[t] \bold{s}_i^H[t]] = \bold{0}, \forall i \neq m$. The diagram of shared scheme is shown in Fig.~\ref{FigShare} and described as follows.

By applying a transmit beamformer $\bold{W}_m \in \mathbb{C}^{N_{tx} \times K}$, the signal $\bold{x}_m[t] \in \mathbb{C}^{N_{tx} \times 1}$ transmitted by the $m$-th sensor can be expressed as
\begin{equation} \label{Eq:TxSha}
\bold{x}_m[t] = \bold{W}_m \bold{s}_m[t].
\end{equation}
Due to the limited transmit power of each sensor, the beamformer design should satisfy the power constraint:
\begin{equation} \label{Eq:PowerSha}
\text{tr}({\bold{W}_m\bold{W}_m}^H) \leq P, \forall m.
\end{equation}
The target reflection signal $\bold{y}_m[t] \in \mathbb{C}^{N_{rx} \times 1}$ received at the $m$-th sensor can be expressed as
\begin{equation} \label{Eq:TrSha}
\bold{y}_m[t] = \bold{G}_{mm} \bold{W}_m \bold{s}_m[t] + \bold{\Omega}_{m}[t] + \bold{n}_r[t],
\end{equation}
where $\bold{\Omega}_{m}[t] = \sum_{i \in \mathcal{M}/\{m\}} \bold{G}_{im} \bold{W}_i \bold{s}_i[t] + \sum_{i \in \mathcal{M}/\{m\}} \bold{Q}_{im} \bold{W}_i \bold{s}_i[t]$ is the interference signal, $\bold{G}_{im} \in \mathbb{C}^{N_{rx} \times N_{tx}}$ and $\bold{Q}_{im} \in \mathbb{C}^{N_{rx} \times N_{tx}}$ denotes the \emph{target response matrix} (TRM) and \emph{direct radar channel} (DRC) between the $i$-th and $m$-th sensors, respectively. $\bold{n}_r[t] \in \mathbb{C}^{N_{rx} \times 1}$ is an \emph{additive white Gaussian noise} (AWGN) vector with distribution $\mathcal{N}_{N_{rx}}(\bold{0},\sigma_r^2\bold{I})$.

According to \cite{bekkerman2006target}, the sufficient statistic matrix $\bold{\hat{Y}}_m \in \mathbb{C}^{N_{rx} \times K}$ can be attained by matched filtering (a.k.a. pulse compression), which is
\begin{align}
\bold{\hat{Y}}_m &= \frac{1}{T}\sum_{t = 1}^T\bold{y}_m[t] \bold{s}_m^H[t] \nonumber\\
& = \frac{1}{T}\sum_{t = 1}^T\sum_{i = 1}^M \bold{G}_{im} \bold{W}_i \bold{s}_i[t] \bold{s}_m^H[t] + \frac{1}{T}\sum_{t = 1}^T\sum_{i \in \mathcal{M}/\{m\}} \bold{Q}_{im} \bold{W}_i \bold{s}_i[t] \bold{s}_m^H[t] + \frac{1}{T}\sum_{t = 1}^T\bold{n}_r[t] \bold{s}_m^H[t]. \label{Eq:SsSha}
\end{align}
According to the law of large-number, when the number of slots $T$ is large, one can get 
\begin{align}
\frac{1}{T}\sum_{t = 1}^T\bold{s}_i[t] \bold{s}_m^H[t] \approx \mathbb{E}_t[\bold{s}_i[t] \bold{s}_m^H[t]] = \bold{0},\\
\frac{1}{T}\sum_{t = 1}^T\bold{s}_m[t] \bold{s}_m^H[t] \approx \mathbb{E}_t[\bold{s}_m[t] \bold{s}_m^H[t]] = \bold{I}.
\end{align}
Therefore, the sufficient statistic matrix can be expressed as
\begin{equation} \label{Eq:SsaSha}
\bold{\hat{Y}}_m = \bold{G}_{mm} \bold{W}_m + \bold{N}_m,
\end{equation}
where $\bold{N}_m = \frac{1}{T}\sum_{t = 1}^T\bold{n}_r[t] \bold{s}_m^H[t] \in \mathbb{C}^{N_{rx} \times K}$ denotes the statistic noise matrix. The distribution of $\bold{N}_m$ is given in the lemma below.

\begin{lemma}[Distribution of $\bold{N}_m$] \label{Distribution} \emph{$\bold{N}_m \sim \mathcal{MN}_{N_{rx} \times K}(\bold{0},\frac{\sigma_r}{\sqrt{T}}\bold{I}_{N_{rx} \times N_{rx}},\frac{\sigma_r}{\sqrt{T}}\bold{I}_{K \times K})$}
\end{lemma}
\proof
See Appendix~\ref{App:Distribution}.
\endproof

The corresponding \emph{probability density function} (PDF) of $\bold{\hat{Y}}_m$ is 
\begin{equation} \label{Eq:PDFSha}
p(\bold{\hat{Y}}_m;\bold{G}_{mm}) = \frac{1}{(2\pi\sigma_r^2/T)^{N_{rx} K/2}}e^{-\frac{T}{2\sigma_r^2}\text{tr}[(\bold{\hat{Y}}_m-\bold{G}_{mm} \bold{W}_m)^H(\bold{\hat{Y}}_m-\bold{G}_{mm} \bold{W}_m)]}.
\end{equation}
Therefore, the MLE of $\bold{G}_{mm}$ can be found by minimizing the log-likelihood function
\begin{equation} \label{Eq:LogSha}
L(\hat{\bold{G}}_{mm}) = \text{tr}[((\bold{\hat{Y}}_m-\bold{G}_{mm} \bold{W}_m)^H(\bold{\hat{Y}}_m-\bold{G}_{mm} \bold{W}_m)].
\end{equation}
The derivatives of $L(\hat{\bold{G}}_{mm})$ w.r.t. $\hat{\bold{G}}_{mm}$ is
\begin{equation} \label{Eq:DerSha}
\frac{\partial L(\hat{\bold{G}}_{mm})}{\partial \hat{\bold{G}}_{mm}} = 2\hat{\bold{G}}_{mm} \bold{W}_m \bold{W}_m^H-2\bold{\hat{Y}}_m \bold{W}_m^H.
\end{equation}
By setting the derivatives of $L(\hat{\bold{G}}_{mm})$ w.r.t. $\hat{\bold{G}}_{mm}$ as zero, the MLE of $\bold{G}_{mm}$ can be obtained as
\begin{equation} \label{Eq:PDFSep}
\hat{\bold{G}}_{mm} = \bold{\hat{Y}}_m \bold{W}_m^H(\bold{W}_m \bold{W}_m^H)^{-1}, \forall m.
\end{equation}
Accordingly, the MSE of estimating $\bold{G}_{mm}$ can be computed as \cite{liu2021integrated}
\begin{equation} \label{Eq:MSEGSha}
\text{MSE}(\bold{G}_{mm}) = \mathbb{E}\left\{\|\bold{G}_{mm} - \bold{\hat{G}}_{mm}\|^2\right\} = \frac{N_{rx}\sigma_r^2}{T}\text{tr}\left\{(\bold{W}_{m}\bold{W}_{m}^H)^{-1}\right\}.
\end{equation}
Given the sensing MSE threshold $\eta_m$, the sensing quality requirement of the $m$-th sensor is
\begin{equation} \label{Eq:QuaSha}
\frac{N_{rx}\sigma_r^2}{T}\text{tr}\left\{(\bold{W}_{m}\bold{W}_{m}^H)^{-1}\right\} \leq \eta_m,~\forall m.
\end{equation}


Due to the long distance between the target and the AP, the target reflection signal vanishes at the AP. Therefore, the corresponding received signal  $\bold{\hat{z}}[t] \in \mathbb{C}^{K \times 1}$ at the AP can be expressed as
\begin{equation} \label{Eq:APSha}
\hat{\bold{z}}[t] = \bold{A}^H \sum_{m=1}^M \bold{H}_{m} \bold{W}_m \bold{s}_m[t]+ \bold{A}^H \bold{n}_c[t],
\end{equation}
where $\bold{H}_{m} \in \mathbb{C}^{N_a \times N_{tx}}$ is the channel between the AP and the $m$-th sensor, $\bold{A} \in \mathbb{C}^{N_a \times K}$ is the data aggregation beamformer at the AP. $\bold{n}_c \in \mathbb{C}^{N_a \times 1}$ is a AWGN vector with distribution $\mathcal{N}_{N_a \times 1}(\bold{0},\sigma_c^2\bold{I})$, which is statistically independent of $\bold{s}_m[t]$. 

Due to the nature of analog transmission, the accuracy of AirComp is prone to the distortion by channel fading and noise. As the goal of AirComp is to accurately compute certain functions, the computation error becomes a natural performance metric. Following the existing literatures (see, e.g., \cite{zhu2018mimo}, \cite{li2019wirelessly}), the error is measured via the MSE between the estimated function value and the ground truth one, i.e.,
\begin{align}
& \mathbb{E}_t\left[\text{tr} \left(\left(\sum_{m=1}^M (\bold{A}^H \bold{H}_{m} \bold{W}_m \!-\! \bold{I})\bold{s}_m[t] \!+\! \bold{A}^H \bold{n}_c[t] \right) \left(\sum_{m=1}^M (\bold{A}^H \bold{H}_{m}\bold{W}_m \!-\! \bold{I})\bold{s}_m[t] \!+\! \bold{A}^H \bold{n}_c[t]\right)^H\right)\right] \nonumber\\
&= \sum_{m=1}^M \text{tr} \left((\bold{A}^H \bold{H}_{m} \bold{W}_m - \bold{I})(\bold{A}^H \bold{H}_{m} \bold{W}_m - \bold{I})^H\right) + \sigma_c^2\text{tr}(\bold{A}\bold{A}^H). \label{MSEASha}
\end{align}

\subsection{Separated Scheme}
\begin{figure*}[ht]
\centering
\includegraphics[scale=0.5]{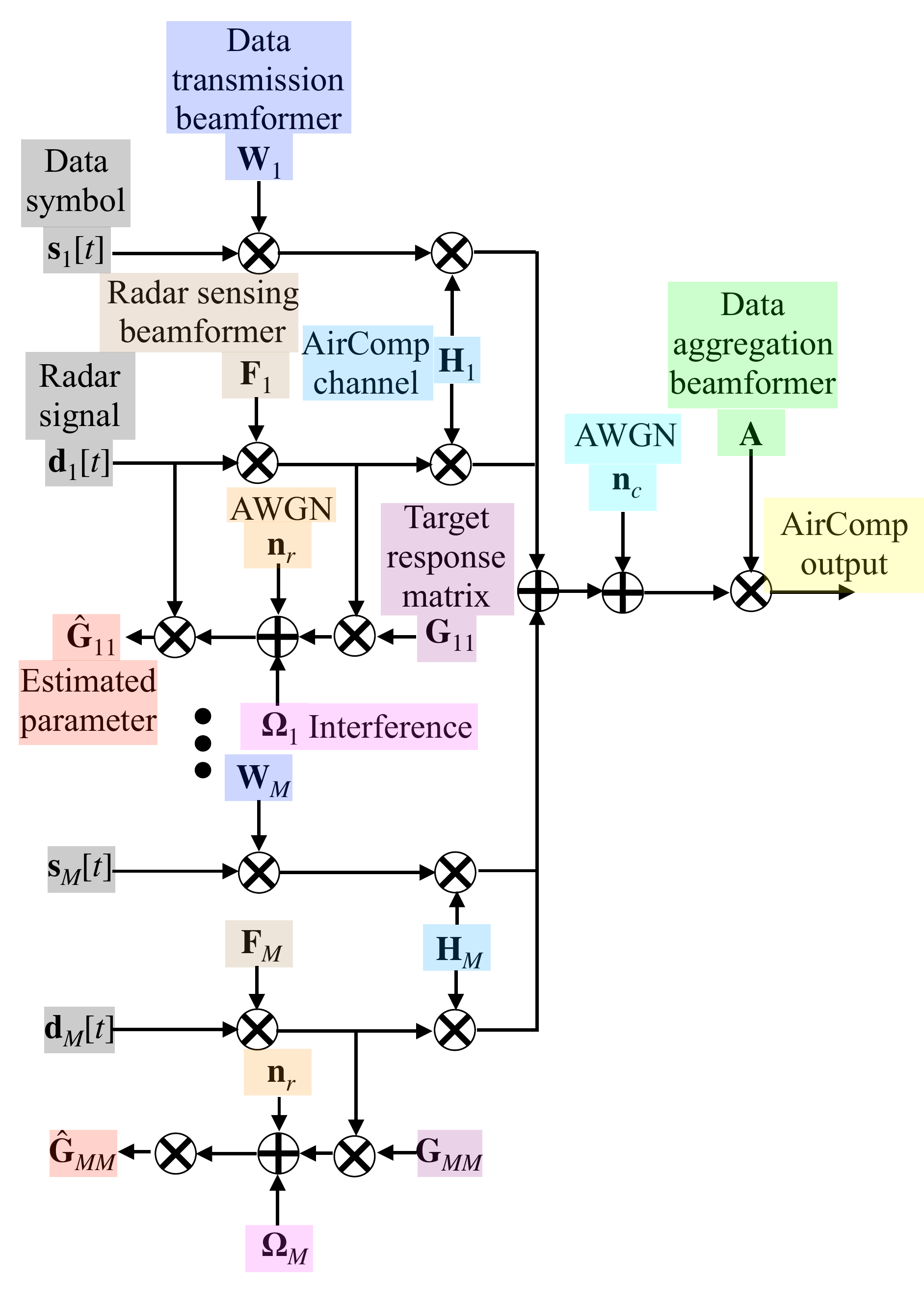}
\caption{Diagram of separated scheme.}
\label{FigSeparated}
\end{figure*}

As shown in Fig. \ref{FigSys} (b), the antennas at each sensor are divided into two groups with $N_s = N_r + N_c$, where $N_r$ antennas are for radar sensing and $N_c$ for data transmission. The $N_r$ antennas are further divided into two groups with $N_r = N_{tx} + N_{rx}$, where $N_{tx}$ antennas are for radar signal transmitting and the remaining $N_{rx}$ for receiving. The data symbols transmitted by the $m$-th sensor at the $t$-th slot can be expressed as a vector denoted by $\bold{d}_m[t] \in \mathbb{C}^{K \times 1}$, where $K$ represents the number of functions to be calculated via AirComp. The data symbols are assumed to be i.i.d. among different sensors and functions with zero mean and unit variance, i.e., $\mathbb{E}_t[\bold{d}_m[t] \bold{d}_m^H[t]] =  \bold{I}$ and $\mathbb{E}_t[\bold{d}_m[t] \bold{d}_i^H[t]] = \bold{0}, \forall i \neq m$. The radar signals transmitted by the $m$-th sensor at the $t$-th slot can be expressed as a vector denoted by $\bold{s}_m[t] \in \mathbb{C}^{K \times 1}$, which satisfies $\mathbb{E}_t[\bold{s}_m[t] \bold{s}_m^H[t]] =  \bold{I}$ and $\mathbb{E}_t[\bold{s}_m[t] \bold{s}_i^H[t]] = \bold{0}, \forall i \neq m$. The data stream signals are statically independent of the radar signals, i.e., $\mathbb{E}_t[\bold{s}_i[t] \bold{d}_m^H[t]] =  \bold{0},\forall i,m$. The diagram of separated scheme is shown in Fig.~\ref{FigSeparated} and described as follows.

The signal transmitted by the $m$-th sensor can be expressed as
\begin{equation} \label{Eq:TxSep}
\bold{x}_m[t] = 
{ \left[ 
\begin{array}{c}
\bold{W}_m \bold{d}_m[t] \\
\bold{F}_m \bold{s}_m[t] \\
\end{array}
\right] },
\end{equation}
where $\bold{W}_m \in \mathbb{C}^{N_c \times K}$ is the data transmission beamformer, $\bold{F}_m \in \mathbb{C}^{N_{tx} \times K}$ is the radar sensing beamformer. Due to the limited transmit power of each sensor, the beamformer design should satisfy the power constraint:
\begin{equation} \label{Eq:PowerSep}
\text{tr}({\bold{W}_m\bold{W}_m}^H) + \text{tr}({\bold{F}_m\bold{F}_m}^H) \leq P, \forall m.
\end{equation}
The target reflection signal $\bold{y}_m[t] \in \mathbb{C}^{N_{rx} \times 1}$ received at the $m$-th sensor can be expressed as
\begin{equation} \label{Eq:TrSep}
\bold{y}_m[t] = \bold{G}_{mm} \bold{F}_m \bold{s}_m[t] + \bold{\Omega}_m[t] + \bold{n}_r[t],
\end{equation}
where $\bold{\Omega}_m[t] = \sum_{i=1}^M \bold{C}_{im} \bold{W}_i \bold{d}_i[t] + \sum_{i \in \mathcal{M}/\{m\}} (\bold{G}_{im} \bold{F}_i \bold{s}_i[t] + \bold{Q}_{im} \bold{F}_i \bold{s}_i[t] + \bold{O}_{im} \bold{W}_i \bold{d}_i[t])$. For the $i$-th and $m$-th sensors, $\bold{G}_{im} \in \mathbb{C}^{N_{rx} \times N_{tx}}$ is the TRM, $\bold{Q}_{im} \in \mathbb{C}^{N_{rx} \times N_{tx}}$ is the DRC, $\bold{C}_{im} \in \mathbb{C}^{N_{rx} \times N_c}$ is the data reflection channel, and $\bold{O}_{im} \in \mathbb{C}^{N_{rx} \times N_c}$ is the direct data channel. $\bold{n}_r[t] \in \mathbb{C}^{N_{rx} \times 1}$ is an AWGN vector with distribution $\mathcal{N}_{N_{rx}}(\bold{0},\sigma_r^2\bold{I})$. 

According to \cite{bekkerman2006target}, the sufficient statistic matrix $\bold{\hat{Y}}_m \in \mathbb{C}^{N_{rx} \times K}$ can be derived as 
\begin{align}
\bold{\hat{Y}}_m &= \frac{1}{T}\sum_{t = 1}^T\sum_{i = 1}^M ( \bold{G}_{im} \bold{F}_i \bold{s}_i[t] + \bold{C}_{im} \bold{W}_i \bold{d}_i[t])\bold{s}_m^H[t] \nonumber\\
& + \frac{1}{T}\sum_{t = 1}^T\sum_{i \in \mathcal{M}/\{m\}} (\bold{Q}_{im} \bold{F}_i \bold{s}_i[t] + \bold{O}_{im} \bold{W}_i \bold{d}_i[t])\bold{s}_m^H[t] + \frac{1}{T}\sum_{t = 1}^T\bold{n}_r[t] \bold{s}_m^H[t]. \label{Eq:SsSep}
\end{align}
According to the law of large-number, when the number of slots $T$ is large, one can get 
\begin{align}
&\frac{1}{T}\sum_{t = 1}^T\bold{s}_i[t] \bold{s}_m^H[t] \approx \mathbb{E}_t[\bold{s}_i[t] \bold{s}_m^H[t]] = \bold{0},\\
&\frac{1}{T}\sum_{t = 1}^T\bold{s}_m[t] \bold{s}_m^H[t] \approx \mathbb{E}_t[\bold{s}_m[t] \bold{s}_m^H[t]] = \bold{I},\\
&\frac{1}{T}\sum_{t = 1}^T\bold{d}_i[t] \bold{s}_m^H[t] \approx \mathbb{E}_t[\bold{d}_i[t] \bold{s}_m^H[t]] = \bold{0}.
\end{align}
Therefore, the sufficient statistic matrix can be expressed as
\begin{equation} \label{Eq:SsaSep}
\bold{\hat{Y}}_m = \bold{G}_{mm} \bold{F}_m + \bold{N}_m,
\end{equation}
where $\bold{N}_m = \frac{1}{T}\sum_{t = 1}^T\bold{n}_s[t] \bold{s}_m^H[t] \in \mathbb{C}^{N_{rx} \times K}$. Following the similar analysis in the shared scheme, it can be derived that $\bold{N}_m \sim \mathcal{MN}_{N_{rx} \times K}(\bold{0},\frac{\sigma_r}{\sqrt{T}}\bold{I}_{N_{rx} \times N_{rx}},\frac{\sigma_r}{\sqrt{T}}\bold{I}_{K \times K})$. 

Accordingly, the MSE of estimating $\bold{G}_{mm}$ can be computed as \cite{liu2021integrated}
\begin{equation} \label{Eq:MSEGSep}
\text{MSE}(\bold{G}_{mm}) = \mathbb{E}\left\{\|\bold{G}_{mm} - \bold{\hat{G}}_{mm}\|^2\right\} = \frac{N_{rx}\sigma_r^2}{T}\text{tr}\left\{(\bold{F}_{m}\bold{F}_{m}^H)^{-1}\right\}.
\end{equation}
Given the sensing MSE threshold $\eta_m$, the sensing quality requirement of the $m$-th sensor is
\begin{equation} \label{Eq:QuaSep}
\frac{N_{rx}\sigma_r^2}{T}\text{tr}\left\{(\bold{F}_{m}\bold{F}_{m}^H)^{-1}\right\} \leq \eta_m,~\forall m.
\end{equation}


The received signal $\bold{\hat{z}}[t] \in \mathbb{C}^{K \times 1}$ at the AP can be expressed as
\begin{equation} \label{Eq:APSep}
\hat{\bold{z}}[t] = \bold{A}^H \sum_{m=1}^M (\bold{H}_{m} \bold{W}_m \bold{d}_m[t]+\bold{R}_{m} \bold{F}_m \bold{s}_m[t]) +\bold{A}^H \bold{n}_c[t],
\end{equation}
where $\bold{H}_{m} \in \mathbb{C}^{N_a \times N_c}$ and $\bold{R}_{m} \in \mathbb{C}^{N_a \times N_{tx}}$ are the channels between the AP and the $m$-th sensor for data symbols and radar signals, respectively. The AWGN vector $\bold{n}_c \in \mathbb{C}^{N_a \times 1}$ is statistically independent of $\bold{s}_m[t]$ and $\bold{d}_m[t]$. The corresponding MSE between the estimated function value and the ground truth one can be expressed as
\begin{align}
& \mathbb{E}_t\left[\left|\bold{A}^H \sum_{m=1}^M (\bold{H}_{m} \bold{W}_m \bold{d}_m[t] + \bold{R}_{m} \bold{F}_m \bold{s}_m[t])+ \bold{A}^H \bold{n}_c[t] - \sum_{m=1}^M \bold{d}_m[t]\right|^2\right] \label{MSEASep}\\
&= \sum_{m=1}^M \text{tr} \left((\bold{A}^H \bold{H}_{m} \bold{W}_m - \bold{I})(\bold{A}^H \bold{H}_{m} \bold{W}_m - \bold{I})^H\right) + \sum_{m=1}^M \text{tr} \left(\bold{A}^H \bold{R}_{m} \bold{F}_{m} \bold{F}_m^H \bold{R}_{m}^H \bold{A}\right) + \sigma_c^2\text{tr}(\bold{A}\bold{A}^H). \nonumber
\end{align}

\section{Dual-functional Shared Beamforming Design}
The shared scheme for radar sensing and AirComp in the ISCCO system can be formulated as a joint optimization problem over transmission beamformer $\bold{W}_m$ at each sensor and aggregation beamformer $\bold{A}$ at the AP. Specifically, given the MSE in \eqref{MSEASha} together with the power constraint in \eqref{Eq:PowerSha} and the sensing quality constraint in \eqref{Eq:QuaSha}, the problem can be formulated as
\begin{equation*}
\begin{aligned}
\min_{\bold{A}, \{\bold{W}_m\}} \quad 
&\sum_{m=1}^M \text{tr} \left((\bold{A}^H \bold{H}_{m} \bold{W}_m - \bold{I})(\bold{A}^H \bold{H}_{m} \bold{W}_m - \bold{I})^H\right) + \sigma_c^2\text{tr}(\bold{A}\bold{A}^H)\\ 
\textbf{(P1)} \qquad \text{s.t.} \qquad 
&\text{tr}\left((\bold{W}_{m}\bold{W}_{m}^H)^{-1}\right) \leq \frac{T\eta_m}{N_{rx}\sigma_r^2}, \forall m,\\
&\text{tr}({\bold{W}_m\bold{W}_m}^H) \leq P, \forall m.
\end{aligned}
\end{equation*}
Problem P1 is difficult to solve due to its \emph{non-convexity}. The lack of convexity arises from the coupling between the transmitting and aggregation beamformers. To deal with such problem, the optimal transmission beamforming design is given in the following proposition. 

\begin{proposition}[Optimal Transmission Beamformer at Sensor]\label{ZFSha} \emph{
For the shared beamforming design, given the data aggregation beamformer $\bold{A}$, the computation error is minimized by adopting the following zero-forcing transmission beamformer at the sensors:
\begin{equation}\label{Eq:ZFSha}
\bold{W}_m = (\bold{H}_m^H\bold{A}\bold{A}^H\bold{H}_m)^{-1} \bold{H}_m^H\bold{A} , \forall m.
\end{equation}
}
\end{proposition}
\proof
See Appendix~\ref{App:ZFSha}.
\endproof

By adopting the zero-forcing transmitting beamformer, the corresponding problem can be formulated as
\begin{equation*}
\begin{aligned}
\min_{\bold{A}} \quad 
& \sigma_c^2\text{tr}(\bold{A}\bold{A}^H)\\ 
\textbf{(P2)} \qquad \text{s.t.} \quad 
&\text{tr}\left(\bold{H}_m^H\bold{A}\bold{A}^H\bold{H}_m\right) \leq \frac{T\eta_m}{N_{rx}\sigma_r^2}, \forall m,\\
&\text{tr}((\bold{H}_m^H\bold{A}\bold{A}^H\bold{H}_m)^{-1}) \leq P, \forall m.
\end{aligned}
\end{equation*}
Since $\text{tr}((\bold{H}_m^H\bold{A}\bold{A}^H\bold{H}_m)^{-1})$ is neither convex nor concave over $\bold{A} \in \mathbb{C}^{N_a \times K}$, the problem (P2) is non-convex. By introducing new variable $\bold{\hat{A}} = \bold{A} \bold{A}^H$, the problem can be formulated as
\begin{equation*}
\begin{aligned}
\min_{\bold{\hat{A}}} \quad 
& \sigma_c^2\text{tr}(\bold{\hat{A}})\\ 
\textbf{(P3)} \qquad \text{s.t.} \quad 
&\text{tr}\left(\bold{H}_m^H\bold{\hat{A}}\bold{H}_m\right) \leq \frac{T\eta_m}{N_{rx}\sigma_r^2}, \forall m,\\
&\text{tr}((\bold{H}_m^H\bold{\hat{A}}\bold{H}_m)^{-1}) \leq P, \forall m,\\
&\text{rank}(\bold{\hat{A}}) = K,\\
&\bold{\hat{A}} \succeq 0.
\end{aligned}
\end{equation*}
By applying the \emph{semidefinite relaxation} (SDR), the problem can be formulated as
\begin{equation*}
\begin{aligned}
\min_{\bold{\hat{A}}} \quad 
& \sigma_c^2\text{tr}(\bold{\hat{A}})\\ 
\textbf{(P4)} \qquad \text{s.t.} \quad 
&\text{tr}\left(\bold{H}_m^H\bold{\hat{A}}\bold{H}_m\right) \leq \frac{T\eta_m}{N_{rx}\sigma_r^2}, \forall m,\\
&\text{tr}((\bold{H}_m^H\bold{\hat{A}}\bold{H}_m)^{-1}) \leq P, \forall m,\\
&\bold{\hat{A}} \succeq 0.
\end{aligned}
\end{equation*}
The convexity of problem (P4) is established in the following lemma.

\begin{lemma} \label{Convexity} \emph{
Problem P4 is a convex problem.}
\end{lemma}
\proof
See Appendix~\ref{App:Convexity}.
\endproof

Upon solving the problem (P4) via a convex problem solver (e.g., the cvx toolbox in MATLAB) and attaining the globally optimal solution $\hat{\bold{A}}^*$, the next task is to retrieve from it a feasible solution to the original problem denoted by $\tilde{\bold{A}}$. Since the rank of $\hat{\bold{A}}^*$ might be larger than $K$, the Gaussian randomization algorithm proposed in \cite{luo2010semidefinite} can be applied to extract $\tilde{\bold{A}}$ from $\hat{\bold{A}}^*$. The main procedure is summarized in Algorithm~\ref{Al:ISCCO}.

\begin{algorithm}[h]
\caption{Gaussian Randomization Algorithm for ISCCO}
\label{Al:ISCCO}
\begin{itemize}
\item{{\bf Initialization}: Given an SDR solution $\hat{\bold{A}}^*$, and the number of random samples $N$.}
\item{{\bf Gaussian Random Sampling}: \\
(1) Perform eigen decomposition $[\bold{V}_{\hat{\bold{A}}}, \bold{\Sigma}_{\hat{\bold{A}}}] = \text{eig}(\hat{\bold{A}}^*)$.\\
(2) Generate $N'$ random matrices $\bold{Z}_n \sim \mathcal{CN}(\bold{0},\bold{I})$ with $\bold{Z}_n \in \mathbb{C}^{N_a \times K}$, $\bold{0} \in \mathbb{C}^{N_a \times K}$ and $\bold{I} \in \mathbb{C}^{N_a \times N_a}$.\\
(3) Retrieve $N$ feasible solutions $\{\bold{A}_{n} = \bold{V}_{\hat{\bold{A}}} \bold{\Sigma}_{\hat{\bold{A}}}^{1/2} \bold{Z}_n^H\}$, such that the constraints $\text{rank}(\bold{A}_n \bold{A}_n^H) = K$, $\text{tr}((\bold{H}_m^H\bold{A}_n \bold{A}_n^H\bold{H}_m)^{-1}) \leq P$, and $\text{tr}\left(\bold{H}_m^H\bold{A}_n \bold{A}_n^H\bold{H}_m\right) \leq \frac{T\eta_m}{N_{rx}\sigma_r^2},~\forall m$ can be enforced.\\
(4) Select the best $\bold{A}_{n}$ that leads to the minimum objective, namely $\bold{A}_{n}^* = \arg \min_{n} \sigma_c^2\text{tr}(\bold{A}_{n}^H\bold{A}_{n})$.\\
(5) Output $\tilde{\bold{A}}=\bold{A}_{n}^*$ as the approximated optimal data aggregation beamformer.}
\end{itemize}
\end{algorithm}

\begin{remark}[Coupling Relationship of Radar Sensing and AirComp in the Shared Scheme]\emph{As shown in problem (P1), one transmission beamformer needs to be designed at each sensor for supporting both radar sensing and AirComp functionalities, which is further correlated to the data aggregation beamformer design at the server via zero-forcing. Therefore, the shared scheme need to guarantee the sensing MSE requirements at the price of sacrificing the AirComp accuracy.}
\end{remark}

\section{Dual-functional Separated Beamforming Design}
In contrast to the shared scheme, the separated scheme should take the joint optimization of data transmission beamformer $\bold{W}_m$, radar sensing beamformer $\bold{F}_m$, and data aggregation beamformer $\bold{A}$ into consideration. Specifically, given the MSE in \eqref{MSEASep} together with the power constraint in \eqref{Eq:PowerSep} and the sensing quality constraint in \eqref{Eq:QuaSep}, the problem can be formulated as

\begin{equation*}
\begin{aligned}
\min_{\bold{A}, \{\bold{W}_m\}, \{\bold{F}_m\}} \quad 
& \sum_{m=1}^M \text{tr} \left((\bold{A}^H \bold{H}_{m} \bold{W}_m - \bold{I})(\bold{A}^H \bold{H}_{m} \bold{W}_m - \bold{I})^H\right) \\
& \qquad + \sum_{m=1}^M \text{tr} \left(\bold{A}^H \bold{R}_{m} \bold{F}_m \bold{F}_m^H \bold{R}_{m}^H \bold{A}\right) + \sigma_c^2\text{tr}(\bold{A}\bold{A}^H) \\ 
\textbf{(P5)} \qquad \text{s.t.} \qquad 
&\text{tr}\left((\bold{F}_{m}\bold{F}_{m}^H)^{-1}\right) \leq \frac{T\eta_m}{N_{rx}\sigma_r^2}, \forall m,\\
&\text{tr}({\bold{W}_m\bold{W}_m}^H) + \text{tr}({\bold{F}_m\bold{F}_m}^H) \leq P, \forall m.
\end{aligned}
\end{equation*}
Following the similar approach of solving problem (P1), the zero-forcing data transmission beamformer is adopted to minimize the AirComp MSE, i.e.,
\begin{equation}\label{Eq:ZFSep}
\bold{W}_m = (\bold{H}_m^H\bold{A}\bold{A}^H\bold{H}_m)^{-1} \bold{H}_m^H\bold{A} , \forall m.
\end{equation}
The corresponding problem can be formulated as
\begin{equation*}
\begin{aligned}
\min_{\bold{A}, \{\bold{F}_m\}} \quad 
& \sum_{m=1}^M \text{tr} \left(\bold{F}_m \bold{F}_m^H \bold{R}_{m}^H \bold{A}\bold{A}^H \bold{R}_{m} \right) + \sigma_c^2\text{tr}(\bold{A}\bold{A}^H) \\ 
\textbf{(P6)} \qquad \text{s.t.} \qquad 
&\text{tr}\left((\bold{F}_{m}\bold{F}_{m}^H)^{-1}\right) \leq \frac{T\eta_m}{N_{rx}\sigma_r^2}, \forall m,\\
&\text{tr}((\bold{H}_m^H\bold{A}\bold{A}^H\bold{H}_m)^{-1}) + \text{tr}({\bold{F}_m\bold{F}_m}^H)\leq P, \forall m.
\end{aligned}
\end{equation*}
The problem (P6) is non-convex due to the coupling variables $\bold{F}_m$ and $\bold{A}$ in the objective function. Following a common approach in the MIMO beamforming literature (see e.g., \cite{love2005limited,choi2006interpolation,peters2010cooperative}), the radar sensing beamformer $\bold{F}_m$ is constrained to be an orthogonal matrix. Mathematically, one can write $\bold{F}_m = \sqrt{\alpha_m} \bold{D}_m$ with $\bold{D}_m$ being a tall unitary matrix and thus $\bold{D}_m\bold{D}_m^H = \bold{I}$, while $\alpha_m$ is a positive scaling factor. The corresponding problem can be formulated as
\begin{equation*}
\begin{aligned}
\min_{\bold{A}, \{\alpha_m\}} \quad 
& \sum_{m=1}^M \alpha_m \text{tr} \left(\bold{R}_{m}^H \bold{A}\bold{A}^H \bold{R}_{m} \right) + \sigma_c^2\text{tr}(\bold{A}\bold{A}^H) \\ 
\textbf{(P7)} \qquad \text{s.t.} \qquad 
&\frac{N_{tx}}{\alpha_m} \leq \frac{T\eta_m}{N_{rx}\sigma_r^2}, \forall m,\\
&\text{tr}((\bold{H}_m^H\bold{A}\bold{A}^H\bold{H}_m)^{-1}) + \alpha_m \leq P, \forall m.
\end{aligned}
\end{equation*}
It can be observed that the increasing of $\alpha_m$ will result in larger MSE. Therefore, the minimum of MSE over $\alpha_m$ is achieved when the minimum $\alpha_m^*$ is adopted for all $m$, i.e.,
\begin{equation} \label{Eq:Optalpha}
\alpha_m^* = \frac{N_{tx}N_{rx}\sigma_r^2}{T\eta_m}, \forall m.
\end{equation}
By introducing $\bold{\hat{A}} = \bold{A} \bold{A}^H$ and applying the SDR, the problem can be formulated as
\begin{equation*}
\begin{aligned}
\min_{\bold{\hat{A}}} \quad 
& \sum_{m=1}^M \frac{N_{tx}N_{rx}\sigma_r^2}{T\eta_m} \text{tr} \left(\bold{R}_{m}^H \bold{\hat{A}} \bold{R}_{m} \right) + \sigma_c^2\text{tr}(\bold{\hat{A}}) \\ 
\textbf{(P8)} \qquad \text{s.t.} \quad 
&\text{tr}((\bold{H}_m^H\bold{\hat{A}}\bold{H}_m)^{-1}) + \frac{N_{tx}N_{rx}\sigma_r^2}{T\eta_m} \leq P, \forall m,\\
&\bold{\hat{A}} \succeq 0.
\end{aligned}
\end{equation*}
It can be easily proved that problem (P8) is convex due to the linear objective and convex constraints. Upon attaining the globally optimal solution of problem (P8) by convex programming, denoted by $\hat{\bold{A}}^*$, the remaining task is to convert it into a feasible solution of the original problem, denoted by $\tilde{\bold{A}}$, of rank $K$. To this end, the Gaussian randomization based Algorithm~\ref{Al:ISCCO} is applied. 

\begin{remark}[Coupling Relationship of Radar Sensing and AirComp in the Separated Scheme]\emph{As shown in problem (P8), the existence of radar signals results in extra AirComp error denoted by $\sum_{m=1}^M \frac{N_{tx}N_{rx}\sigma_r^2}{T\eta_m} \text{tr} \left(\bold{R}_{m}^H \bold{\hat{A}} \bold{R}_{m} \right)$. To mitigate the interference on AirComp caused by radar signals, the radar sensing beamformers are designed to achieve the maximum tolerance of sensing MSE, i.e., $\frac{N_{rx}\sigma_r^2}{T} \text{tr}\left((\bold{F}_{m}\bold{F}_{m}^H)^{-1}\right) = \eta_m$.}
\end{remark}


\begin{figure*}[ht]
\centering
\includegraphics[scale=0.45]{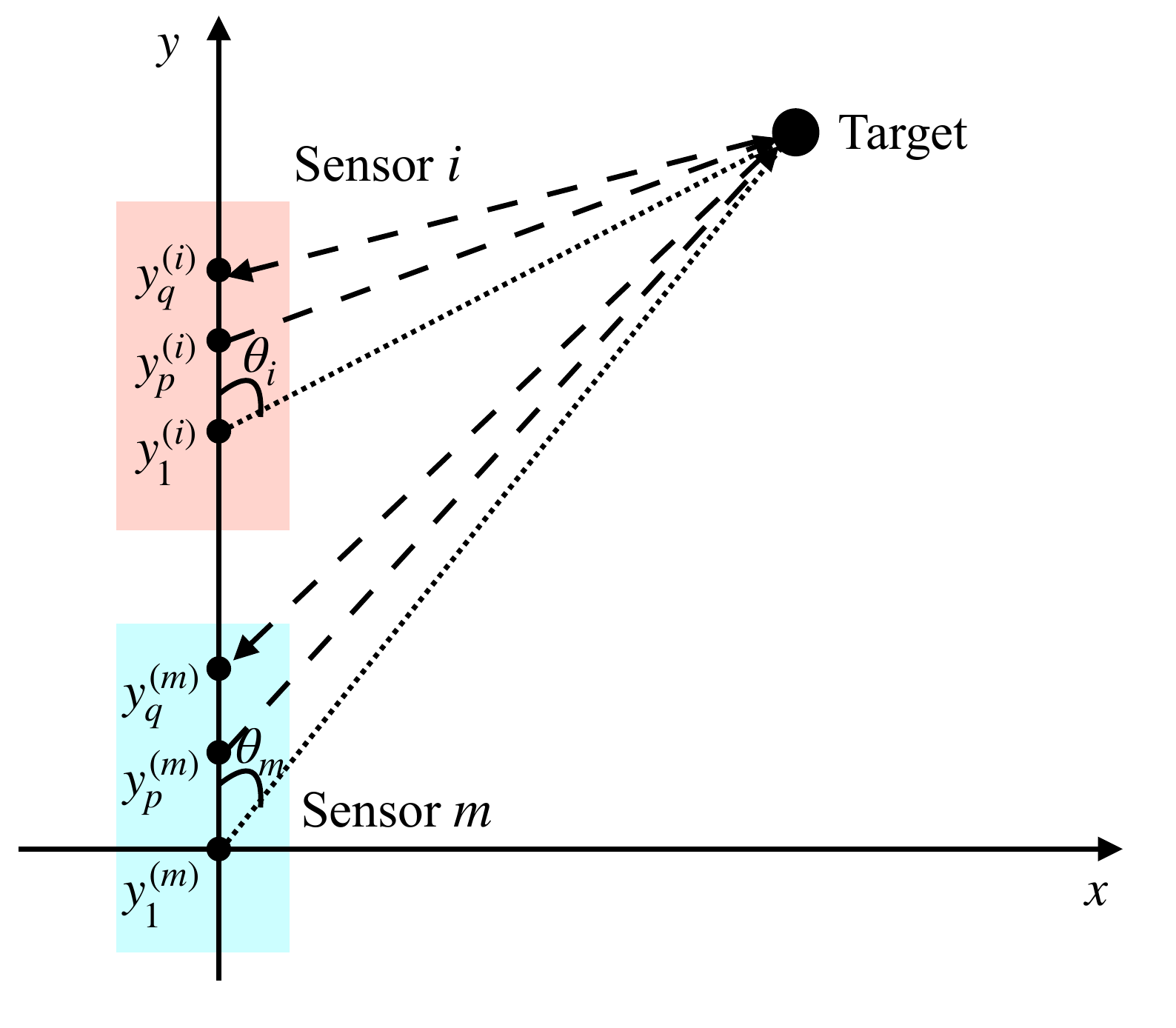}
\caption{Location Estimation System based on ISCCO.}
\label{FigLocation}
\end{figure*}

\section{Target Location Estimation based on ISCCO}
In this section, the ISCCO scheme was applied for the use case of target location estimation. Particularly, the location of the target is estimated by all $M$ sensors based on their own locations as well as the information of angle and distance extracted from the reflected radar signals. The estimated location of the target by each sensor is then transmitted to the server via AirComp, and thus the server will obtain the averaged estimated target location. As shown in Fig.~\ref{FigLocation}, the TRM $\bold{G}_{mm}$ is composed of a phase delay matrix $\bold{\Phi}(\theta_m)$ and a complex amplitude $\beta_{m}$ of the received signal, i.e., $\bold{G}_{mm} = \beta_{m} \bold{\Phi}(\theta_m)$. Let $\phi_{pq}(\theta_m)$ denote the element of $\bold{\Phi}(\theta_m)$ in $p$-th row and $q$-th column, then
\begin{equation} \label{Eq:PDM}
\phi_{pq}(\theta_m) = \exp\{-j\omega[\tau_p(\theta_m)+\tau_q(\theta_m)]\},
\end{equation}
where $\omega$ represents the angular velocity, $\tau_p(\theta_m)$ represents the transmitting time delay between the $1$-st and $p$-th antennas, $\tau_q(\theta_m)$ represents the receiving time delay between the $1$-st and $q$-th antennas. According to \cite{bekkerman2006target}, the phase delay between the $p$-th and $q$-th antennas at the $m$-th sensor can be expressed as
\begin{equation} \label{Eq:PD}
\phi_{pq}(\theta_m) = \exp\{-\frac{2\pi j}{\lambda}(y_p^{(m)}+y_q^{(m)})\sin{\theta}_m\},
\end{equation}
where $y_p^{(m)}$ and $y_q^{(m)}$ denote the location of the $p$-th and $q$-th antennas at the $m$-th sensor, respectively. Following the derivation of $\bold{G}_{mm}$, the MLE of $\beta_{m}$ and $\theta_m$ can be found by minimizing the log-likelihood function:
\begin{equation} \label{Eq:MLELoc}
L(\beta_{m},\theta_m) = \text{tr}[(\bold{\hat{Y}}_m - \beta_{m} \bold{\Phi}(\theta_m) \bold{W}_m)^H(\bold{\hat{Y}}_m - \beta_{m} \bold{\Phi}(\theta_m) \bold{W}_m)].
\end{equation}
The derivatives of $L(\beta_{m},\theta_m)$ w.r.t. $\beta_{m}$ is
\begin{equation} \label{Eq:Derbeta}
\frac{\partial L(\beta_{m},\theta_m)}{\partial \beta_{m}} = 2 \beta_{m} \text{tr}(\bold{W}_m^H \bold{\Phi}^H(\theta_m) \bold{\Phi}(\theta_m) \bold{W}_m) - 2\text{tr} (\bold{W}_m^H \bold{\Phi}^H(\theta_m) \bold{\hat{Y}}_m).
\end{equation}
According to \eqref{Eq:PDFSep}, $\hat{\bold{G}}_{mm} \bold{W}_m \bold{W}_m^H = \bold{\hat{Y}}_m \bold{W}_m^H, \forall m$. Setting the derivatives as zero, one can get
\begin{equation} \label{Eq:Estbeta}
\hat{\beta}_m = \frac{\text{tr} (\bold{W}_m^H \bold{\Phi}^H(\theta_m) \hat{\bold{G}}_{mm} \bold{W}_m)}{\text{tr}(\bold{W}_m^H \bold{\Phi}^H(\theta_m) \bold{\Phi}(\theta_m) \bold{W}_m)}, \forall m.
\end{equation}
By replacing $\beta_{m}$ with $\hat{\beta}_m$ in $L(\beta_{m},\theta_m)$, one can get
\begin{equation} \label{Eq:Estphi}
L(\theta_m)= \text{tr}(\bold{\hat{Y}}_m^H\bold{\hat{Y}}_m)-\frac{\text{tr}^2(\bold{W}_m^H\bold{\Phi}^H(\theta_m)\hat{\bold{G}}_{mm} \bold{W}_m)}{\text{tr}(\bold{W}_m^H\bold{\Phi}^H(\theta_m)\bold{\Phi}(\theta_m)\bold{W}_m)}.
\end{equation}
As the angle $\theta_m$ to be estimated is only relevant with the second item in \eqref{Eq:Estphi}, one can get
\begin{equation} \label{Eq:Esttheta}
\hat{\theta}_m = \arg\max_{\theta_m} \frac{\text{tr}^2(\bold{W}_m^H\bold{\Phi}^H(\theta_m)\hat{\bold{G}}_{mm} \bold{W}_m)}{\text{tr}(\bold{W}_m^H\bold{\Phi}^H(\theta_m)\bold{\Phi}(\theta_m)\bold{W}_m)}.
\end{equation}
It should be noted that $\hat{\theta}_m$ cannot be expressed in closed form. Therefore, the grid search or golden section search can be applied to find the numerical results, where the beamformer $\bold{W}_m$ is obtained by solving problem (P1). On the other hand, the distance $d_m$ between the target and the $m$-th sensor can be estimated following the free space propagation law \cite{sturm2011waveform}. Based on the estimated parameters (distance $\hat{d}_m$ and angle $\hat{\theta}_m$) and its own location $(0,y_m)$, the $m$-th sensor can obtain its local estimation of the target location denoted by $\bold{\hat{z}}_m = [\hat{x}_m,\hat{y}_m]^T$ via
\begin{align} 
\hat{x}_m &= \hat{d}_m \sin \hat{\theta}_m, \label{Eq:Estx} \\ 
\hat{y}_m &= y_1^{(m)} + \hat{d}_m \cos \hat{\theta}_m. \label{Eq:Esty}
\end{align}

The target location estimated by the $m$-th sensor is then modulated into data symbols represented by $\bold{s}_m = [x_m,y_m]^T$, where
\begin{align} 
x_m &= \frac{\hat{x}_m}{\bar{x}} - 1, \label{Eq:Modx} \\ 
y_m &= \frac{\hat{y}_m}{\bar{y}} - 1, \label{Eq:Mody}
\end{align}
with $\bar{x}$ and $\bar{y}$ denoting the statistic values of the target location at x-axis and y-axis. After transmission beamforming $\bold{W}_m$, the data symbols are transmitted to the AP. In the shared scheme, the signals received at the AP can be expressed as 
\begin{equation} \label{Eq:RecSha}
\hat{\bold{s}} = \bold{A}^H \sum_{m=1}^M \bold{H}_{m} \bold{W}_m \bold{s}_m+ \bold{A}^H \bold{n}_c,
\end{equation}
where $\hat{\bold{s}} = [x',y']^T$. The averaged estimated target location can be derived as $\bold{z}_m = [(x' +1)\bar{x},(y' + 1)\bar{y}]^T$. The performance of radar sensing and AirComp will be evaluated based on the simulation results in section VII.

\section{Simulation}
In this section, the performance of our proposed ISCCO framework is evaluated by simulation, where the radar sensing and AirComp channel models in shared and separated schemes are simulated based on \eqref{Eq:TrSha}, \eqref{Eq:APSha}, \eqref{Eq:TrSep}, and \eqref{Eq:APSep}. The performance metric is the normalized AirComp MSE, defined by $\text{MSE}/M$ with the AirComp MSE given in \eqref{MSEASha} and \eqref{MSEASep} for shared and separated schemes, respectively. The simulation parameters are set as follows unless specified otherwise. The number of time slots is $T = 1000$. The number of computed functions is set to be $K=10$. There are $M=10$ sensors each equipped with $N_s = 12$ antennas and one AP with $N_a = 15$ antennas. In the shared scheme, $N_{tx} = 6$ antennas are for signal transmitting and $N_{rx} = 6$ antennas are for signal receiving. In the separated scheme, $N_{c} = 4$ antennas are for data transmission and $N_{r} = 8$ antennas are for radar sensing, where $N_{tx} = 4$ antennas are for radar signal transmitting and $N_{rx} = 4$ antennas are for radar signal receiving. All the channels are assumed to be i.i.d. \emph{Rician fading}, modeled as i.i.d. complex Gaussian random variables with non-zero mean $\mu = 1$ and variance $\sigma^2 = 1$. In addition, the maximum transmission power is set as $P_0 = 10$ mW. The effective power conversion efficiency follows a uniform distribution with $\eta_n \in (0,1)$. According to the settings in LTE \cite{schwarz2013lte}, the powers of noise in radar signal channel $\sigma_r^2$ and data transmission channel $\sigma_c^2$ are $-79.5$ dBm. Each point in the figures is obtained by averaging over 10 simulation realizations, with independent channels in each realization.

\subsection{Baseline Schemes}
Two baselines are designed by applying \emph{antenna selection} (AS) on the shared and separated schemes, respectively. All the schemes assume the channel-inversion data precoding. Define the sum-channel matrix $\bold{H}_{\sf sum} = \sum_{m=1}^M \bold{H}_m$. The baseline schemes with AS select the $K$ receive antenna observing the largest channel gains in the sum channel $\bold{H}_{\sf sum}$. For fair comparison, all aggregation beamformers in the baseline schemes are scaled to have the same norm.

\subsection{Function Computation Performance of ISCCO}
\begin{figure}[t]
\centering
\includegraphics[scale=0.5]{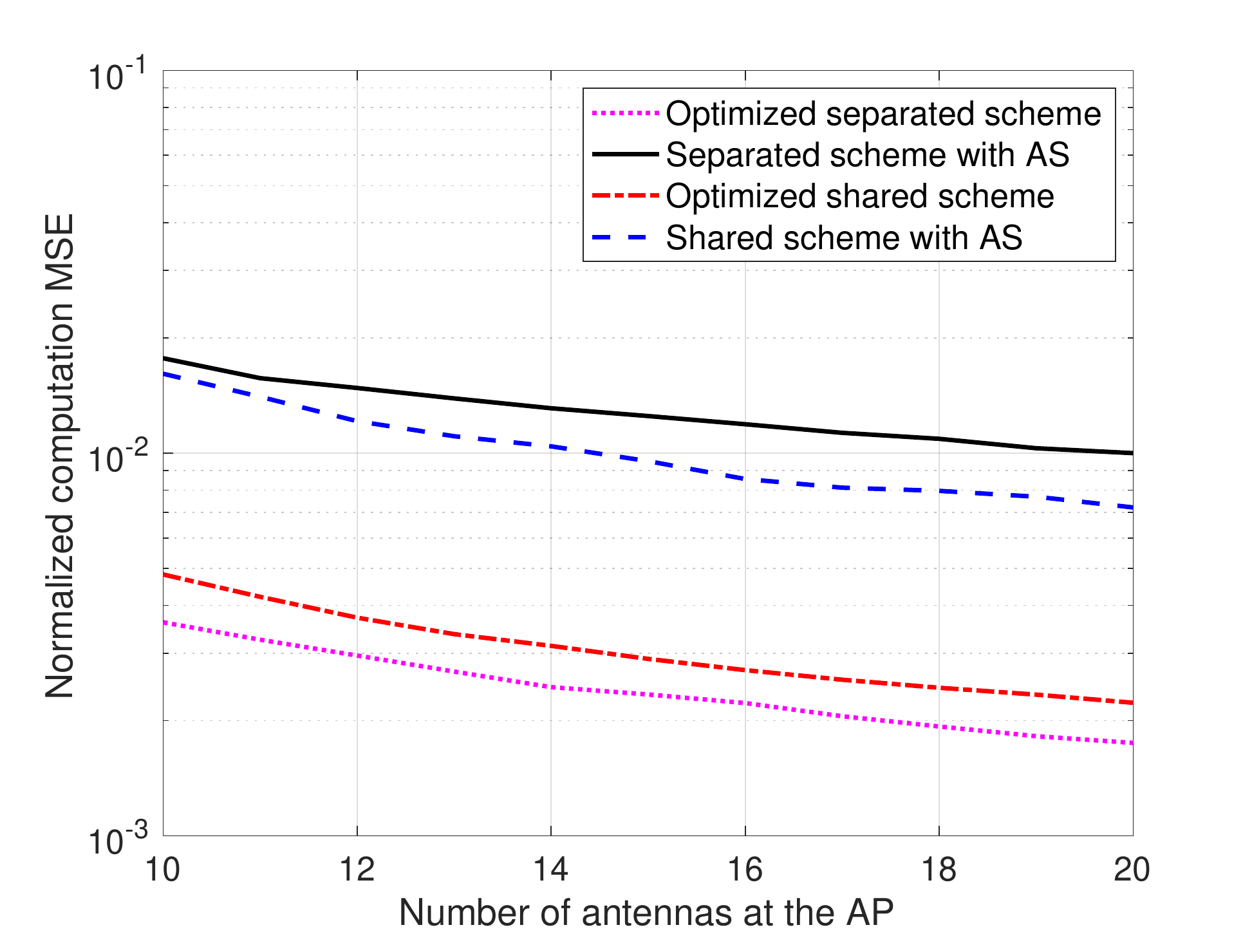}
\caption{Normalized AirComp MSE versus the number of antennas at the AP.}
\label{FigNaAircomp}
\end{figure}

First, the normalized AirComp MSE versus the number of antennas at the AP is evaluated in Fig.~\ref{FigNaAircomp} for both the shared and separated schemes. It can be observed that the normalized AirComp MSE decreases with the increasing number of antennas at the AP. This is because more antennas at the AP will enlarge the dimension of data aggregation beamformer, and thus exploit the diversity gain for achieving lower AirComp MSE. It should be noted that both the shared and separated schemes proposed in this paper can achieve lower AirComp MSE than the baselines with AS, which verifies the necessity of beamformer optimization. Moreover, the separated scheme has better performance than the shared one under the current system settings. The reason is that the dual-functional signals in the shared scheme make it hard to design one common beamformer for supporting both radar sensing and AirComp, while the interference caused by radar signals in the separated scheme can be effectively mitigated by the dedicated beamformer design for AirComp signals.

\begin{figure}[t]
\centering
\includegraphics[scale=0.5]{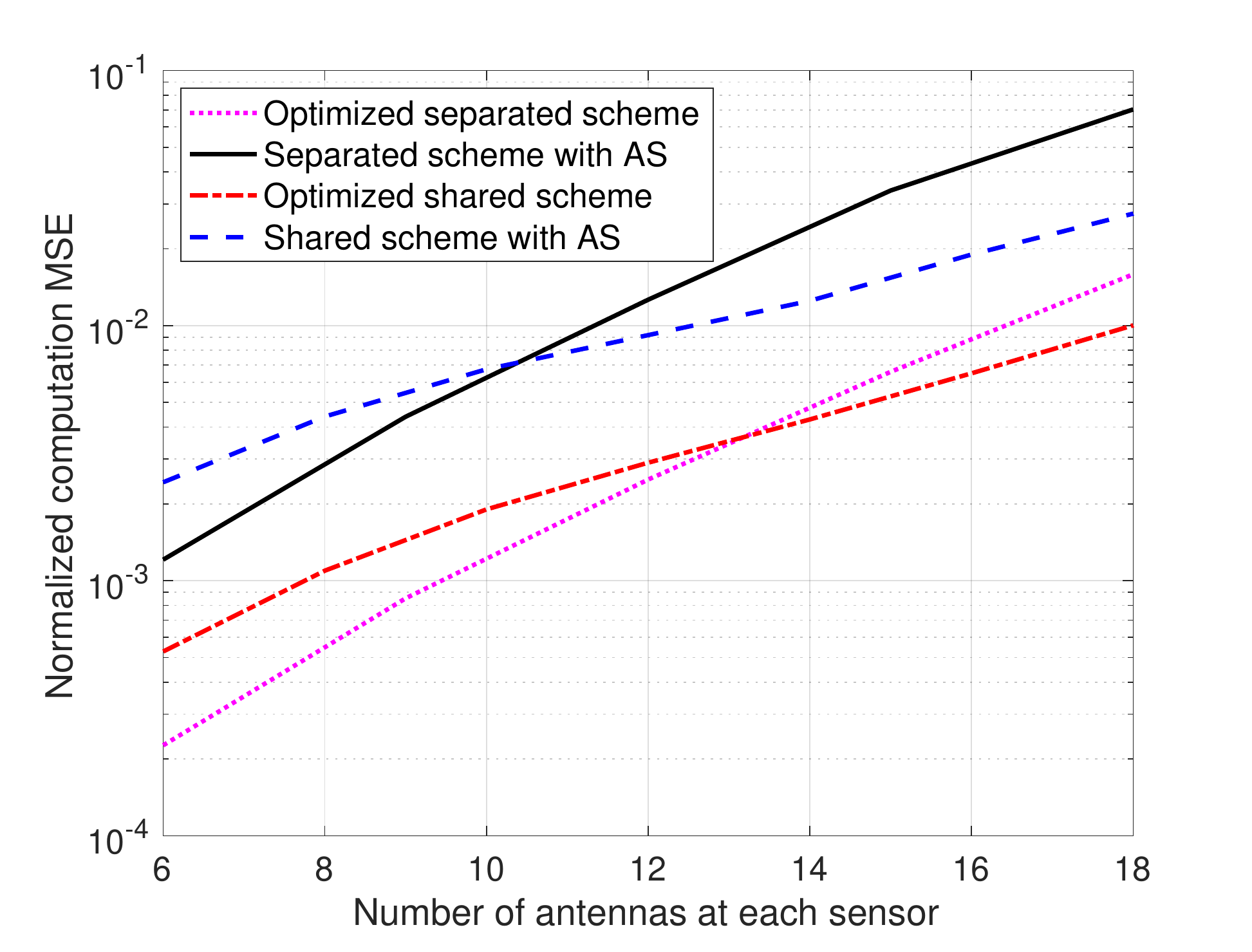}
\caption{Normalized AirComp MSE versus the number of antennas at each sensor.}
\label{FigNsAircomp}
\end{figure}

Fig.~\ref{FigNsAircomp} demonstrates the normalized AirComp MSE versus the number of antennas at each sensor in both the shared and separated schemes. One can observe that the normalized AirComp MSE monotonically increases with the increasing number of antennas at each sensor, since more antennas at sensors will result in larger dimension of TRM to be estimated and thus more stringent sensing constraints. Therefore, the beamformers need to guarantee the requirements of radar sensing at the price of scarifying the performance of AirComp. Moreover, the performance of shared scheme becomes better than that of the separated one with the increasing number of antennas at each sensor. Such a phenomenon is caused by double effects of deploying more antennas at each sensor. On one hand, more antennas at each sensor will result in larger dimension of beamforming matrix for supporting the dual-functionality of signals in the shared scheme. On the other hand, more antennas for radar sensing at each sensor will exacerbate the interference on AirComp in the separated scheme. The similar trends also hold for the baselines with AS.

\begin{figure}[t]
\centering
\includegraphics[scale=0.5]{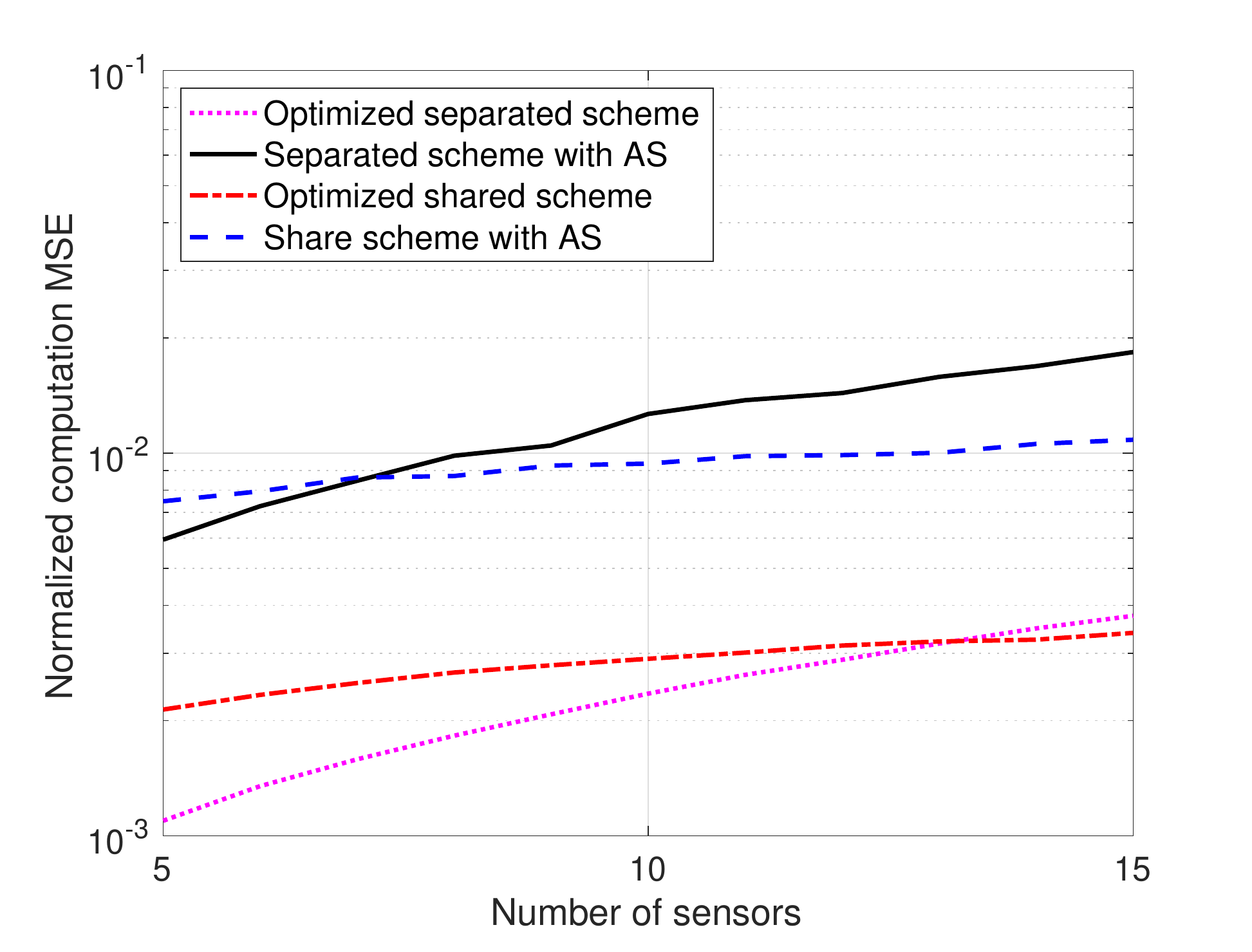}
\caption{Normalized AirComp MSE versus the number of sensors.}
\label{FigMAircomp}
\end{figure}

Fig.~\ref{FigMAircomp} illustrates the curves of the normalized AirComp MSE versus the number of sensors for both the shared and separated schemes. It is shown that the increasing number of sensors will result in higher normalized computation MSE, as more connected sensors make it harder to design one common data aggregation beamformer to equalize the channels among different sensors. Moreover, the increasing trend of normalized AirComp MSE in the separated scheme is more drastic compared with the shared scheme, since larger number of sensors will exacerbate the interference of radar signals on AirComp. The similar trends also  hold for the baselines with AS.

\begin{figure}[t]
\centering
\includegraphics[scale=0.5]{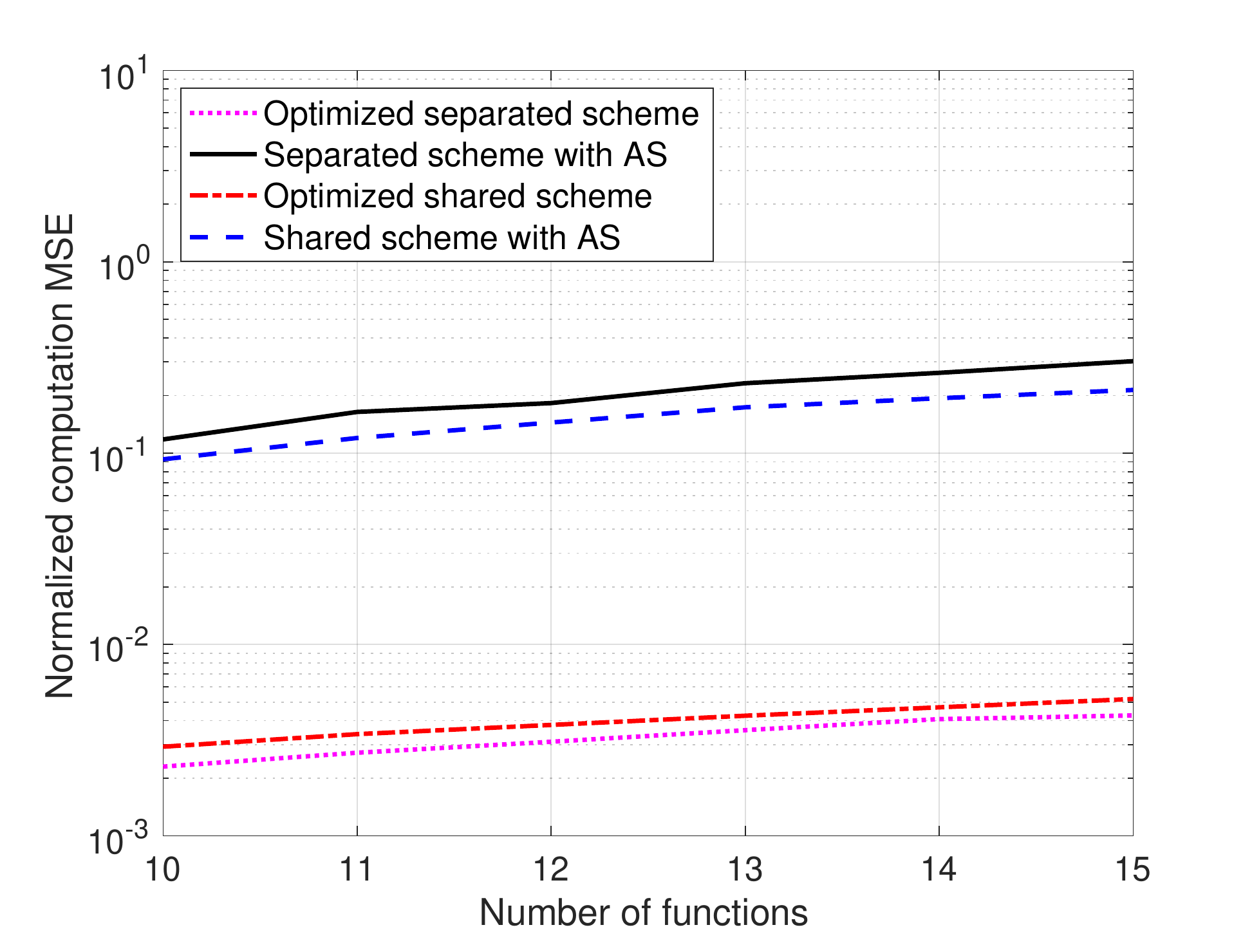}
\caption{Normalized AirComp MSE versus the number of functions to be computed.}
\label{FigKAircomp}
\end{figure}

Fig.~\ref{FigKAircomp} further shows the curves of the normalized AirComp MSE versus the number of functions to be computed in both the shared and separated schemes. One can observe that the normalized AirComp MSE increases with the number of  functions, which indicates that higher computation throughput is at a cost of declining accuracy. Moreover, the separated scheme always performs better than the shared one no matter how many functions need to be computed, which implies that the former is more robust against the varying number of functions. The similar trends also hold for the baselines with AS.

\subsection{Radar Sensing Performance of ISCCO}
\begin{figure}[t]
\centering
\includegraphics[scale=0.5]{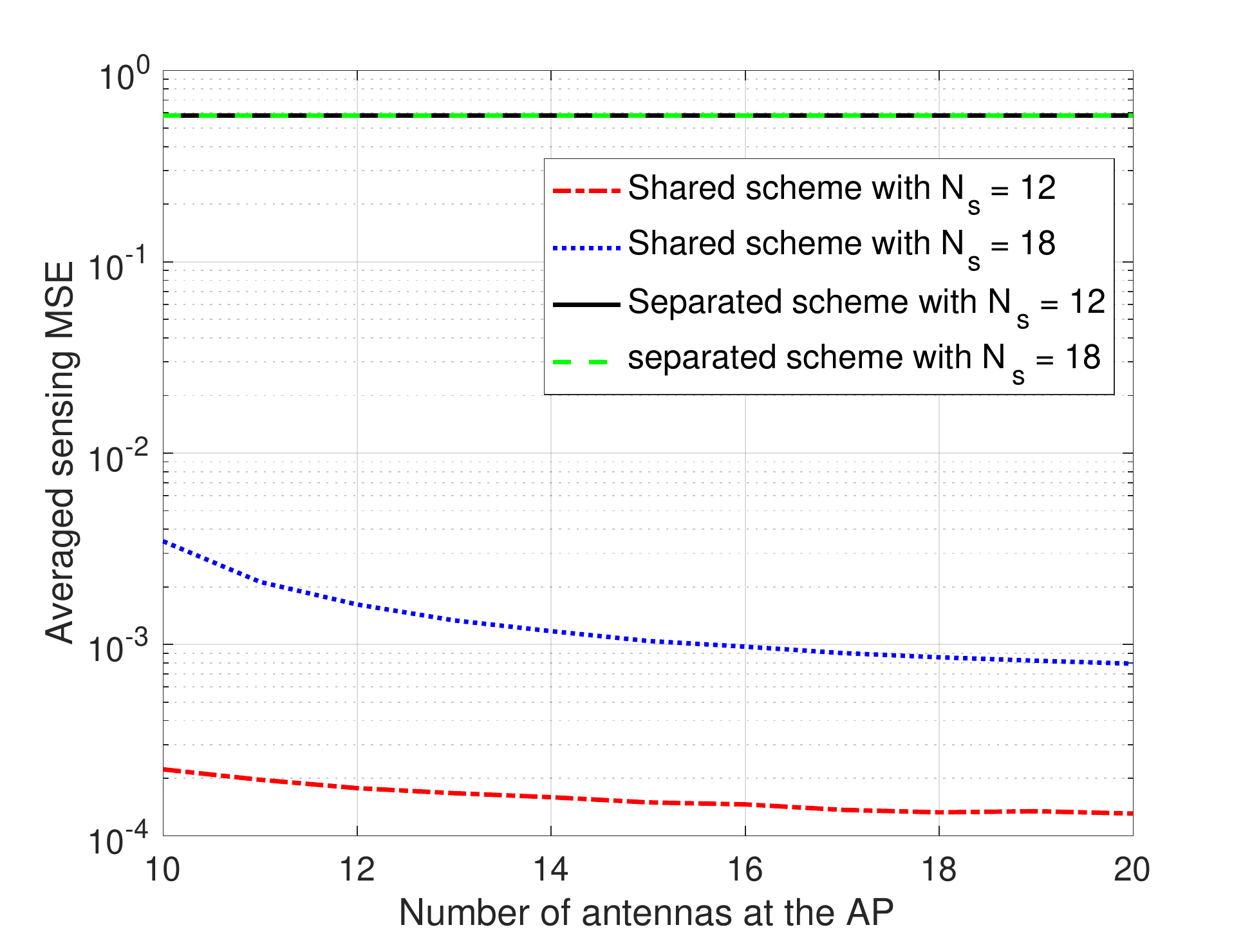}
\caption{Effects of antenna amounts on averaged sensing MSE.}
\label{FigSense}
\end{figure}

As for radar sensing, the effects of antenna amounts at both the sensors and the AP on the averaged sensing MSE are illustrated in Fig.~\ref{FigSense} for both the shared and separated schemes. One can observe that the averaged sensing MSE decreases with the increasing number of antennas at the AP in the shared scheme, which indicates that the enlarged dimension of aggregation beamformer will result in higher degree of freedom for achieving lower sensing MSE. Moreover, deploying more antennas at each sensor in the shared scheme will result in larger averaged sensing MSE as the dimension of TRM to be estimated is enlarged. In contrast, the averaged sensing MSE does not change with the number of antennas at neither the AP nor the sensors in the separated scheme, since the radar sensing constraint is tighten for mitigating the interference of radar signals on AirComp. Therefore, the sensing MSE only depends on the sensing quality requirement and is irrelevant with other parameters.

\subsection{Target Location Estimation based on ISCCO}
\begin{figure}[hh]
\centering
\includegraphics[scale=0.5]{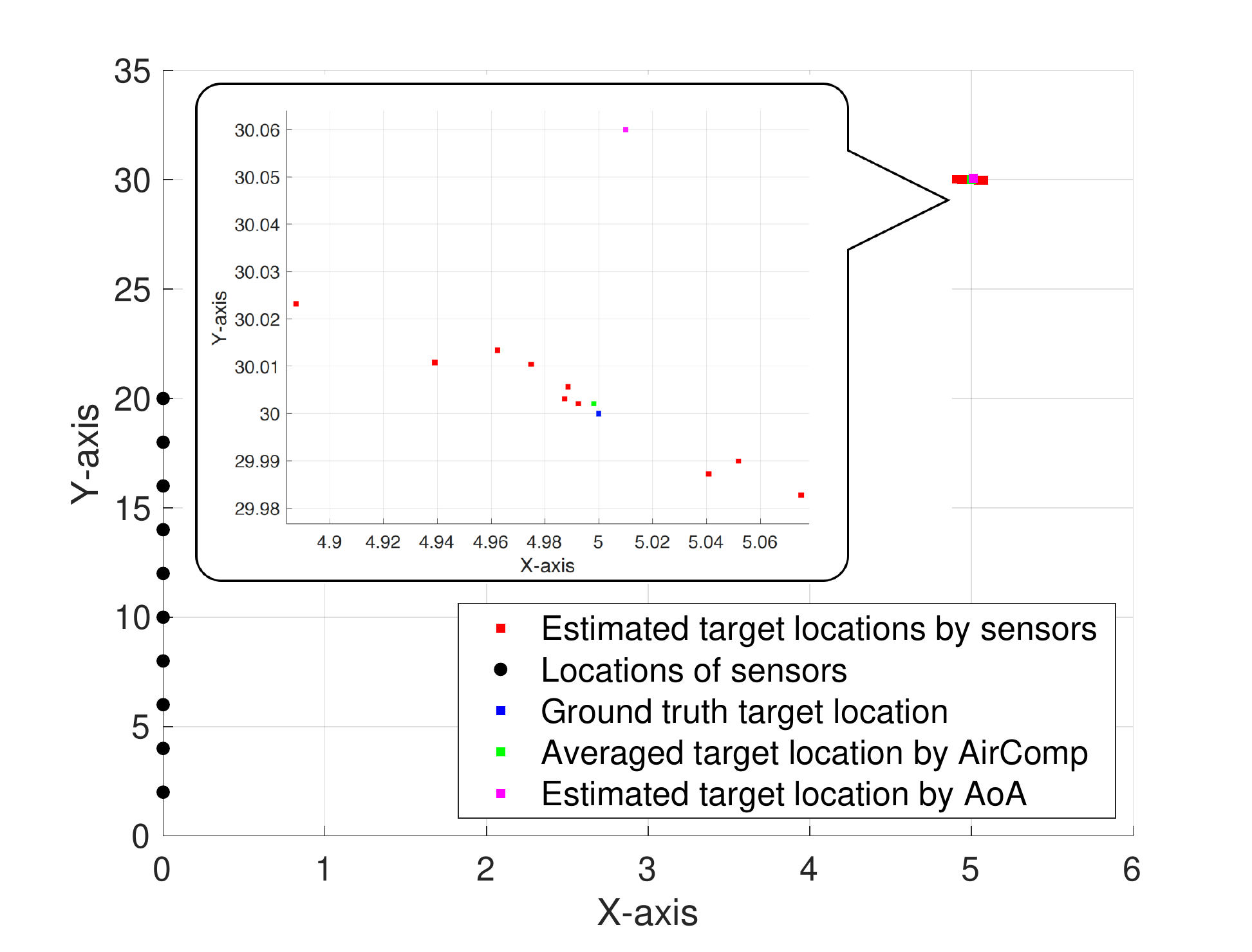}
\caption{Target location estimation based on ISCCO.}
\label{FigTarget}
\end{figure}

The use case of target location estimation based on ISCCO is demonstrated in Fig. \ref{FigTarget}. The ground truth location of the target is set as $(5,30)$ m. $M = 10$ sensor are located at the range $[0,20]$ m on the Y-axis with $2$ m distance between each other. The number of antennas at each sensor is set as $N_{tx} = N_{rx} = 2$ with $0.1$ m space between each other. The information to be estimated and transmitted is a vector which contains the two-dimensional location of the target. The angles between the target and the sensors are estimated via \eqref{Eq:Esttheta}, while the distance is assumed to be perfectly estimated. The performance of the conventional radar sensing scheme based on \emph{angle of arriving} (AoA) \cite{liu2020joint} is also plotted, where the estimated target location $(x_0,y_0)$ is obtained by minimizing the MSE function $\min_{x_0,y_0} \sum_{m=1}^M |\hat{\theta}_m-\arctan\frac{x_0-x_m}{y_0-y_m}|^2$ via grid search, with $(x_m,y_m)$ denoting the location of the $m$-th sensor. It can be observed that the estimated target location by each sensor based on ISAC is a little deviating from the ground truth, while the application of AirComp can alleviate such deviation by averaging the measured values of sensors over transmission. Moreover, the target location estimated by ISCCO is more approaching the ground truth than that by the conventional AoA.

\begin{figure}[hh]
\centering
\includegraphics[scale=0.5]{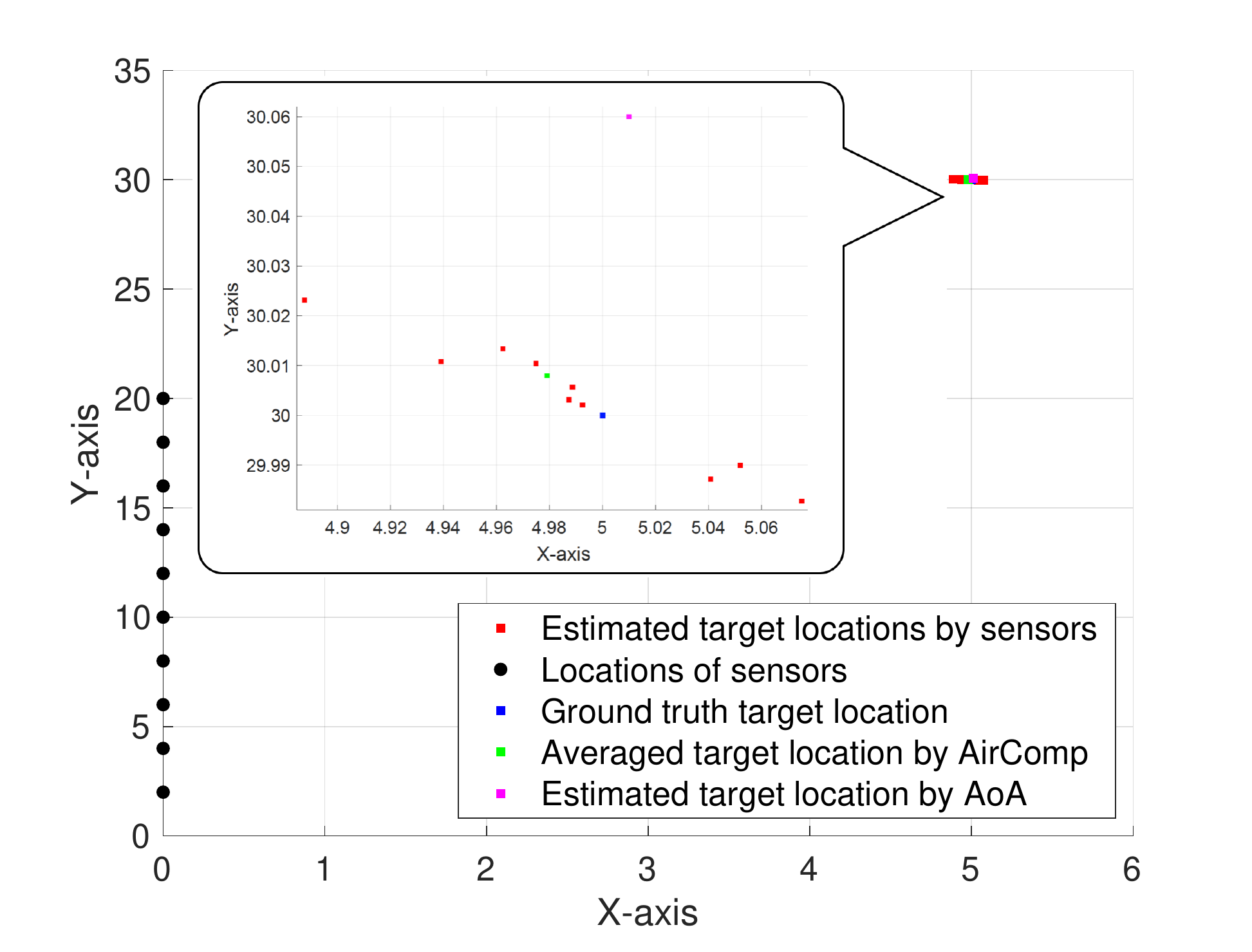}
\caption{Target location estimation based on ISCCO under strong channel noise.}
\label{FigTargetH}
\end{figure}

When the power of noise in data transmission channel is set as $-59.5$ dBm, the target location estimation based on ISCCO is demonstrated in Fig. \ref{FigTargetH}. It can be observed that the performance of AirComp is deteriorated due to the strong noise. In such condition, the target location estimation by a single sensor might have better performance, which necessitates the scheduling of sensors.

\section{Concluding Remarks}
In this paper, an ISCCO framework has beed proposed for enabling the simultaneous radar sensing and Aircomp to improve the spectrum efficiency in IoT systems. To this end, two designs known as the shared and separated schemes have been investigated. In the shared scheme, all the antennas at each sensor are exploited for transceiving dual-functional signals. In the separated scheme, the antenna array at each sensor is divided into two sub-arrays for supporting radar sensing and Aircomp, respectively. The non-convex problem of joint optimizing the beamformers for radar sensing, data transmission and aggregation is solved via semidefinite relaxation together with Gaussian randomization. This work contributes to the promising new research area of ISCCO and many interesting follow-up research issues warrant further investigation, such as sensor scheduling, vehicular tracking, and target surface estimation.


\appendix
\subsection{Proof of Lemma \ref{Distribution}}\label{App:Distribution}
By letting $\bold{S}_m = [\bold{s}_m[1],\bold{s}_m[2],...\bold{s}_m[T]] \in \mathbb{C}^{K \times T}$ and $\bold{N}_r = [\bold{n}_r[1],\bold{n}_r[2],...,\bold{n}_r[T]] \in \mathbb{C}^{N_{rx} \times T}$, one can get $\bold{N}_m = \frac{1}{T}\bold{N}_r \bold{S}_m^H \in \mathbb{C}^{N_{rx} \times K}$. The vectorization of $\bold{N}_r$ is a Gaussian random vector denoted by $\text{vec}(\bold{N}_r) \sim \mathcal{N}_{N_{rx} T \times 1}(\bold{0},\sigma_r^2\bold{I}_{N_{rx} T\times N_{rx} T})$. Correspondingly, the vectorization of $\bold{N}_m$ can be expressed as
\begin{equation}\label{app:vect}
\bold{n}_m = \text{vec}(\bold{N}_m) = \text{vec}(\frac{1}{T}\bold{I}_{N_{rx} \times N_{rx}}\bold{N}_r\bold{S}_m^H) = \frac{1}{T}(\bold{S}_m \otimes \bold{I}_{N_{rx} \times N_{rx}})\text{vec}(\bold{N}_r),
\end{equation}
which is a linear transformation of $\text{vec}(\bold{N}_r)$. Therefore, $\bold{n}_m \sim \mathcal{N}_{N_{rx} K \times 1}(\bold{0},\bold{\Sigma})$, where $\bold{\Sigma} = \mathbb{E}\left[\frac{\sigma_r^2}{T^2}(\bold{S}_m \otimes \bold{I}_{N_{rx} \times N_{rx}})(\bold{S}_m \otimes \bold{I}_{N_{rx} \times N_{rx}})^H\right] = \mathbb{E}\left[\frac{\sigma_r^2}{T^2}(\bold{S}_m \bold{S}_m^H) \otimes \bold{I}_{N_{rx} \times N_{rx}}\right] = \frac{\sigma_r^2}{T}\bold{I}_{N_{rx} K \times N_{rx} K}$. Under such condition the PDF of $\bold{n}_m$ is
\begin{align}\label{app:PDFn}
p(\bold{n}_m|\bold{0},\bold{\Sigma}) & = \frac{1}{(2\pi)^{N_{rx} K/2}|\bold{\Sigma}|^{1/2}}e^{-\frac{1}{2}\bold{n}_m^H \bold{\Sigma}^{-1}\bold{n}_m} \nonumber \\
& = \frac{1}{(2\pi)^{N_r K/2}|\frac{\sigma_r^2}{T}\bold{I}_{N_{rx} K\times N_{rx} K}|^{1/2}}e^{-\frac{T}{2\sigma_r^2}\bold{n}_m^H\bold{n}_m} \nonumber \\
& = \frac{1}{(2\pi)^{N_{rx} K/2} |\frac{\sigma_r}{\sqrt{T}}\bold{I}_{N_{rx} \times N_{rx}}|^{K/2} |\frac{\sigma_r}{\sqrt{T}}\bold{I}_{K \times K}|^{N_{rx}/2}}e^{-\frac{T}{2\sigma_r^2}\text{vec}(\bold{N}_m)^H \text{vec}(\bold{N}_m)} \nonumber\\
& = \frac{1}{(2\pi)^{N_{rx} K/2} |\frac{\sigma_r}{\sqrt{T}}\bold{I}_{N_{rx} \times N_{rx}}|^{K/2}|\frac{\sigma_r}{\sqrt{T}}\bold{I}_{K \times K}|^{N_{rx}/2}}e^{-\frac{T}{2\sigma_r^2}\text{tr}[\bold{N}_m^H\bold{N}_m]} \nonumber\\
& = p(\bold{N}_m|\bold{0},\frac{\sigma_r}{\sqrt{T}}\bold{I}_{N_{rx} \times N_{rx}},\frac{\sigma_r}{\sqrt{T}}\bold{I}_{K \times K}).
\end{align}

\subsection{Proof of Proposition \ref{ZFSha}}\label{App:ZFSha}
Given the AirComp MSE minimization objective provided in \eqref{MSEASha}, it can be observed that both $\sum_{m=1}^M \text{tr} \left((\bold{A}^H \bold{H}_{m} \bold{W}_m - \bold{I})(\bold{A}^H \bold{H}_{m} \bold{W}_m - \bold{I})^H\right) $ and $\sigma_c^2\text{tr}(\bold{A}\bold{A}^H)$ are positive. Therefore, given any data aggregation beamformer $\bold{A}$, the inequality
\begin{equation}\label{Eq:Inequ}
\sum_{m=1}^M \text{tr} \left((\bold{A}^H \bold{H}_{m} \bold{W}_m - \bold{I})(\bold{A}^H \bold{H}_{m} \bold{W}_m - \bold{I})^H\right) + \sigma_c^2\text{tr}(\bold{A}\bold{A}^H) \geq \sigma_c^2\text{tr}(\bold{A}\bold{A}^H)
\end{equation}
always holds. It is easy to verify that setting $\bold{W}_m$ to have the zero-forcing structure in \eqref{Eq:ZFSha} enforces
\begin{equation}\label{Eq:equ}
\sum_{m=1}^M \text{tr} \left((\bold{A}^H \bold{H}_{m} \bold{W}_m - \bold{I})(\bold{A}^H \bold{H}_{m} \bold{W}_m - \bold{I})^H\right)  = 0,
\end{equation}
and thus achieves the equality in \eqref{Eq:Inequ}.

\subsection{Proof of Lemma \ref{Convexity}}\label{App:Convexity}
Since the item $(\bold{H}_m^H\bold{\hat{A}}\bold{H}_m)^{-1}$ is convex over $\bold{\hat{A}}$ and $\text{tr}(\bold{X})$ is linear over $\bold{X}$, the function $\text{tr}((\bold{H}_m^H\bold{\hat{A}}\bold{H}_m)^{-1})$ is convex over $\bold{\hat{A}}$ according to the composition rule \cite{Boyd2006convex}. Since other constraints as well as the objective function are linear functions over $\bold{\hat{A}}$, problem (P4) is convex.

\bibliographystyle{IEEEtran}

\end{document}